\documentclass[conference]{IEEEtran}

\usepackage{soul}

\usepackage[table,dvipsnames]{xcolor}

\usepackage{cite}

\ifCLASSINFOpdf
\else
\fi
\usepackage{amsmath}
\usepackage{multirow}

\usepackage{array}

\ifCLASSOPTIONcompsoc
 \usepackage[caption=false,font=normalsize,labelfont=sf,textfont=sf]{subfig}
\else
 \usepackage[caption=false,font=footnotesize]{subfig}
\fi

\usepackage{xurl}

\usepackage{newtxtext}

\usepackage{url}

\usepackage{enumitem}
\usepackage{amsfonts}
\usepackage{booktabs}
\usepackage{makecell}
\usepackage{siunitx}
\usepackage{microtype}

\usepackage{fix-cm}

\usepackage{sansmath}

\usepackage{marvosym}
\usepackage{bm}
\usepackage{tikz}
\usetikzlibrary{arrows.meta, positioning, fit, backgrounds, calc}
\tikzset{panel/.style={draw, thick, rounded corners,color=red}, private/.style={fill=gray!20, draw=black!50}}
\usepackage{pifont}
\usepackage{algorithm}
\usepackage{algpseudocode}
\usepackage{balance}
\usepackage{placeins}
\usepackage{orcidlink}
\usepackage{tabularx}
\usepackage{hyperref}
\usepackage{circledsteps}
\usepackage{silence}
\usepackage{nicematrix}
\usepackage[most]{tcolorbox}
\newcommand*\nogap[1]{}

\definecolor{ieeeblue}{RGB}{26, 12, 171}
\definecolor{snet-red}{rgb}{0.772, 0.059, 0.122}
\definecolor{sectionColor}{HTML}{B20100}
\definecolor{berlinblue}{HTML}{005A84}
\colorlet{summaryrule}{snet-red!70}
\colorlet{summarybackground}{black!3}

\newsavebox{\summaryboxsave}

\newlength{\shadowfillgap}

\NiceMatrixOptions{
  custom-line={
    command=hdashedline,
    ccommand=cdashedline,
    tikz={
      dash pattern=on 2.5pt off 1.5pt,
      line width=\lightrulewidth
    }
  }
}

\NewDocumentCommand{\negcircled}{O{8pt} O{black} m}{%
  \tikz[baseline=-0.75ex]{
    \node[
      circle,
      fill=#2,
      draw=#2,
      text=white,
      inner sep=0pt,
      minimum size=0.9em,
      font=\fontsize{#1}{#1}\selectfont\bfseries
    ] {#3};
  }%
  \hspace{0.2em}\ignorespaces%
}

\newcommand{\circNum}[1]{%
  \tikz[baseline=(char.base)]{%
    \node[circle,draw=black,text=black,fill=none,inner sep=0.5pt,
      minimum size=1em,line width=0.6pt,font=\small\bfseries] (char) {#1};%
  }%
}
\newcommand{\circRom}[1]{%
  \tikz[baseline=(char.base)]{%
    \node[circle,draw=berlinblue,text=berlinblue,fill=none,inner sep=0.5pt,
      minimum size=1em,line width=0.6pt,font=\small\bfseries] (char) {#1};%
  }%
}

\hypersetup{
    colorlinks=true,
    linkcolor=sectionColor,
    citecolor=ieeeblue,
    filecolor=ieeeblue,
    urlcolor=ieeeblue,
    pdftitle={ShadowPath: Lookup-Private Credential Status Verification over Authenticated State},
    pdfauthor={Anonymous Authors},
    pdfsubject={Anonymous NDSS 2027 review submission},
    pdfkeywords={anonymous credential status, zero-knowledge, revocation, authenticated state}
}


\begin{document}
%
\title{\textsc{ShadowPath}: Lookup-Private Credential Status Verification over Authenticated State}


\author{\IEEEauthorblockN{
Patrick Herbke\kern1pt\orcidlink{0000-0001-9649-2975}\kern0.7pt\textsuperscript{\href{mailto:p.herbke@tu-berlin.de}{\Letter}\IEEEauthorrefmark{2}},
Wolf Rieder\kern1pt\orcidlink{0009-0001-4932-9814}\IEEEauthorrefmark{2},
Christian René Sechting\kern1pt\orcidlink{0009-0008-8869-3396}\IEEEauthorrefmark{2}, \\
Huaning Yang\kern1pt\orcidlink{0009-0000-1041-8484}\IEEEauthorrefmark{4}, 
Sid Lamichhane\kern1pt\orcidlink{0009-0000-5845-7060}\IEEEauthorrefmark{2}, 
Philip Raschke\kern1pt\orcidlink{0000-0002-6738-7137}\IEEEauthorrefmark{2} and
Axel Küpper\kern1pt\orcidlink{0000-0002-4356-5613}\IEEEauthorrefmark{2}}
\IEEEauthorblockA{\IEEEauthorrefmark{2}Technische Universität Berlin, \IEEEauthorrefmark{4}Humboldt-Universität zu Berlin 
}
}


\IEEEoverridecommandlockouts
\makeatletter\def\@IEEEpubidpullup{6.5\baselineskip}\makeatother
\IEEEpubid{\parbox{\columnwidth}{
		Network and Distributed System Security (NDSS) Symposium 2027\\
		22--26 March 2027, Seoul, Republic of Korea\\
		ISBN 978-1-970672-09-1\\  
		https://dx.doi.org/10.14722/ndss.2027.[23$|$24]xxxx\\
		www.ndss-symposium.org
}
\hspace{\columnsep}\makebox[\columnwidth]{}}

\maketitle



\begin{abstract}
Verifiable credentials let holders present digitally signed claims without requiring the issuer to participate in every presentation. Revocation complicates this privacy model because a verifier must determine whether a credential remains valid. Existing status checks may expose recurring identifiers, registry positions, or request metadata. Such information can serve as stable handles to link separate presentations. \textsc{ShadowPath} moves the credential status lookup to the holder. For each presentation, the holder proves, in zero-knowledge, that the credential has not been revoked under the verifier-selected registry root. The verifier learns the status result but not observable metadata. To the best of our knowledge, we provide the first evaluation of Verkle trees for credential revocation and compare them with sparse Merkle trees to assess their applicability in real world applications. The comparison tests whether reducing path depth with Verkle trees offsets the higher cost of KZG-based authentication. Across 30 desktop trials, median Groth16 proving took \qty{371.6}{\milli\second} with sparse Merkle and \qty{2.11}{\second} with Verkle. Verification took \qty{3.70}{\milli\second} and \qty{7.55}{\milli\second}, respectively. Groth16 Verkle proving took about \qty{3}{\second} on both primary mobile devices. The results show that shorter authenticated paths do not necessarily yield cheaper zero-knowledge proofs. With fresh session randomness, verifier-visible status data do not reveal whether two presentations use the same credential under the stated assumption of session-value independence. This guarantee excludes issuer--verifier collusion and synchronization traffic. 
\end{abstract}


%
\IEEEpeerreviewmaketitle

\section{Introduction}\label{sec:intro}
Credential-status verification must enforce revocation without making private presentations linkable. Decentralized identity supports issuer-independent presentation, limiting what issuers learn about credential usage~\cite{10.1145/3446983.3446992,sedlmeir2021digital}. Status checks weaken this boundary because verifiers need fresh revocation information~\cite{11264637,lueks2017fast,10.1145/3167132.3167303}. Retrieval or verification may expose identifiers, deterministic registry positions, request paths, or timing patterns~\cite{11264637}. We call such recurring metadata a \emph{stable handle}: an anchor allowing verifiers, issuers, or network observers to link presentations of the same credential.

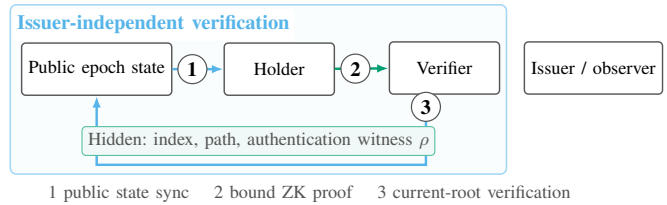
\begin{figure}[t]
   \centering
   \definecolor{spblue}{HTML}{56B4E9}\definecolor{spred}{HTML}{A61C1C}
\definecolor{spgreen}{HTML}{009E73}\definecolor{spgray}{HTML}{555555}
\resizebox{\columnwidth}{!}{%
\begin{tikzpicture}[font=\small, box/.style={draw=spgray,rounded corners=2pt,line width=.7pt,
   minimum width=19mm,minimum height=9mm,align=center,fill=white}, num/.style={circle,draw=spgray,fill=white,line width=.6pt,minimum size=5mm,
   inner sep=0pt,font=\bfseries}, normal/.style={-{Latex[length=2mm]},draw=spgray,line width=1.1pt}, leak/.style={-{Latex[length=2mm]},draw=spred,line width=1.2pt,dashed}, secure/.style={-{Latex[length=2mm]},draw=spblue,line width=1.2pt}]
\path[use as bounding box] (-0.1,-1.82) rectangle (11.45,5.25);
\node[anchor=west,font=\bfseries] at (0,5.0) {Conventional credential-specific lookup};
\node[box] (h1) at (1,4.0) {Holder}; \node[box] (v1) at (3.75,4) {Verifier};
\node[box] (s1) at (6.75,4) {Status registry}; \node[box] (i1) at (10.08,4) {Issuer / observer};
\draw[normal] (h1)--node[num, pos=0.37]{1}(v1); \draw[leak] (v1)--node[num, pos=0.32]{2}(s1); \draw[leak] (s1)--node[num, pos=0.42]{3}(i1);
\node[text=spgray] at (5.5,3.20) {1 presentation \quad 2 credential-specific status query \quad 3 observable request};
\node[draw=spred!65,fill=spred!5,text=spgray,rounded corners=2pt,line width=.55pt,inner sep=3pt] at (5.71,2.67)
 {Stable handles: identifier, registry position, and request metadata};
\draw[black!20,line width=.6pt] (0,2.28)--(11,2.28);
\node[anchor=west,font=\bfseries] at (0,1.95) {\textsc{ShadowPath}: lookup-private status verification};
\draw[spblue!55,fill=spblue!3,rounded corners=3pt,line width=.75pt]
  (0.01,-1.23) rectangle (8.57,1.69);
\node[anchor=west,text=spblue,font=\bfseries,fill=white,inner xsep=2pt] at (0.05,1.35)
  {Issuer-independent verification};
\node[box] (root) at (1.5,0.58) {Public epoch state}; \node[box] (h2) at (4.65,0.58) {Holder};
\node[box] (v2) at (7.5,0.6) {Verifier}; \node[box] (i2) at (10.07,0.6) {Issuer / observer};
\draw[secure] (root)--node[num, pos=0.39]{1}(h2);
\draw[-{Latex[length=2mm]},draw=spgreen,line width=1.2pt] (h2)--node[num, pos=0.38]{2}(v2);
\draw[secure] (7.18,-0.07) |- node[num, at start]{3} (7.18,-1.07) -|(root)  ; 
\node[draw=spgreen!65,fill=spgreen!6,text=spgray,rounded corners=2pt,line width=.55pt,inner sep=3pt] at (4.29,-0.66)
 {Hidden: index, path, authentication witness $\rho$}; 
\node[text=spgray] at (5.17,-1.56) {1 public state sync \quad 2 bound ZK proof \quad 3 current-root verification};
\end{tikzpicture}%
}
   \vspace{-2.5mm}
   \caption{Conventional credential-specific lookup versus \textsc{ShadowPath}'s holder-local status verification. Credential-specific status traffic exposes recurring identifiers and request metadata. \textsc{ShadowPath} derives status locally from shared epoch state and sends a ZKP to the verifier.}   
   \label{fig:stable-handles}
\end{figure}

A traveler may present the same credential for mobility and hotel check-in while disclosing different claims. Figure~\ref{fig:stable-handles} illustrates how credential-status verification can create a link between these presentations. In a conventional credential-specific lookup, the verifier queries status information associated with the presented credential. The lookup may expose an identifier, registry position, or request metadata that recurs across presentations~\cite{11264637,9895264}. \textsc{ShadowPath} shifts credential-specific status lookup from the verifier to the holder. The holder derives the credential status locally from public revocation state. For each presentation, the holder sends a zero-knowledge proof (ZKP) that binds the status result to the credential presentation. The verifier checks the proof against the selected epoch root without learning the registry index or traversal path.

Existing revocation mechanisms make different privacy trade-offs across the status workflow. Online responders expose presentation-time status requests~\cite{rfc6960,10196098}, while status lists reveal registry positions to the verifier~\cite{W3C_Bitstring2024,5751342}. Accumulator-based schemes shift credential-specific work to witness maintenance~\cite{camenisch2002dynamic,10.1145/3634737.3637641}, and Merkle-based zero-knowledge systems hide tree membership within broader credential proofs~\cite{zk-creds,9230363}. \textsc{ShadowPath} addresses the privacy problem across both state acquisition and presentation. It separates credential-independent state acquisition from credential-specific verification, keeps the registry lookup at the holder, authenticates the hidden lookup against the verifier-selected epoch root, and binds the status result to the credential presented in the same session. Sparse Merkle and Verkle backends realize the same protocol with different path-authentication mechanisms. Their matched comparison tests whether shorter authentication paths offset higher per-level authentication cost. 
\newpage
We address the following research questions:
\begin{enumerate}[ label=\textit{RQ\arabic*:}, widest=RQ2:, leftmargin=2.7em, labelsep=0.5em, itemsep=0.2mm, parsep=0pt]

\item How can a holder prove non-revocation from authenticated registry state while hiding credential-specific lookup information? (\autoref{sec:threat-model}--\autoref{sec:problem}, \autoref{sec:implementation}--\autoref{sec:experiments})

\item What proving, verification, and state-distribution costs arise from hiding credential-status lookups? (\autoref{sec:methodology}--\autoref{sec:experiments})
\end{enumerate}

Our contributions are as follows:
\begin{itemize}[itemsep=0.25mm, parsep=0pt]

\item \textit{Stable-handle leakage analysis.}\
Recurring credential-status information is characterized as stable handles and mapped to observation points at the issuer, verifier, and network layers~\cite{W3C_Bitstring2024,10.1145/3634737.3637641,zk-creds}
(\autoref{sec:threat-model}).

\item \textit{Lookup-private credential-status protocol.}\
\textsc{ShadowPath} separates credential-independent acquisition of authenticated epoch state from credential-specific status verification. The holder reconstructs the required witness locally and proves non-revocation against the verifier-selected epoch root without exposing the registry lookup. The same hidden credential identifier binds the status result to the credential presented in the current session. No credential-specific status query leaves the holder during presentation (\autoref{sec:problem}, \autoref{sec:implementation}).

\item \textit{Authenticated-registry design and lifecycle evaluation.}\
Sparse Merkle and Verkle backends realize the same \textsc{ShadowPath} protocol with different path-authentication mechanisms. Their matched evaluation quantifies proving, verification, state synchronization, reconstruction, and mobile costs and exposes the trade-off between authentication-path depth and per-level authentication cost (\autoref{sec:implementation}--\autoref{sec:experiments}).

\end{itemize}
\section{Preliminaries}\label{sec:preliminaries}
Credential-status verification determines whether a presented Verifiable Credential (VC) has been revoked~\cite{sedlmeir2021digital}. The registry evolves through immutable epochs~$e$, each authenticated by a public root~$R_e$. Holders obtain the corresponding state before presentation, while verifiers use~$R_e$ to verify status. Each credential maps to a private registry index~$\mathsf{idx}$ and fingerprint~$\mathsf{fp}$. Non-revoked positions remain empty. Upon revocation, the issuer stores~$\mathsf{fp}$ at the corresponding position.

\subsection{Authenticated Registry Trees}
An authenticated tree organizes registry entries in a tree and cryptographically binds the complete registry state to a root value. Each registry position corresponds to a leaf, while internal nodes authenticate the subtrees below them. Changing a registry entry changes the authentication values along its path and ultimately the root. In \textsc{ShadowPath}, an empty registry position establishes non-revocation. If the position contains a fingerprint, the credential is revoked if the stored fingerprint matches $fp$. We evaluate \textsc{ShadowPath} with two authenticated registry structures: a sparse Merkle tree (SMT) and a Verkle tree. Both authenticate a private registry position against the epoch root~$R_e$ with different path-authentication mechanisms. 

\noindent\textbf{Sparse Merkle tree.}
An SMT is a binary authenticated tree in which each parent node is computed by hashing its two children~\cite{dahlberg2016efficient}. The authentication path for one registry position contains the leaf value and one sibling hash at each tree level. Starting from the leaf, the sibling hashes are used successively to recompute the parent nodes and then the root. A recomputed root equal to~$R_e$ authenticates the registry entry as part of the selected epoch state. For an address space of size~$N$, the authentication path contains $O(\log_2 N)$ sibling hashes.

\noindent\textbf{Verkle tree.}
Compared with the binary SMT, a Verkle tree uses a branching factor~$k$~\cite{kuszmaul2019verkle}. Each internal node contains~$k$ child positions whose values are bound by a compact cryptographic commitment. The holder authenticates the selected child value against the node commitment and repeats this step until reaching the credential's registry position. The resulting sequence of authenticated child selections links the registry position to the epoch root~$R_e$. At the selected registry position, the holder establishes non-revocation from an empty entry or from a stored fingerprint that differs from~$\mathsf{fp}$. In \textsc{ShadowPath}, the holder performs these checks privately inside a ZKP. The verifier checks the resulting proof against~$R_e$, verifying non-revocation without learning the registry index, the traversed path, or the authentication data. A Verkle tree covering~$N$ registry positions has $O(\log_k N)$ levels.

Our Verkle backend uses Kate--Zaverucha--Goldberg (KZG) polynomial commitments~\cite{kate2010constant}. A KZG commitment compactly binds all child values of a Verkle node. For the child position selected by the registry index, the holder provides a KZG opening proving that the claimed child value is the value committed at that position. One opening is required for each traversed tree level. KZG commitments and openings have constant size, but verification requires elliptic-curve pairings and a structured reference string (SRS), a set of public parameters generated during setup. The SMT and Verkle backends expose the authentication trade-off evaluated in this work: the SMT uses more computationally inexpensive hash-based levels, whereas the Verkle tree uses fewer but more expensive KZG-authenticated levels.

\subsection{Zero-Knowledge Proofs}
A ZKP allows a prover to convince a verifier that a statement is true without revealing the private information used to prove it. Formally, the prover shows that a public statement~$x$ and a private witness~$w$ satisfy a relation~$\mathcal{R}$, i.e., $\mathcal{R}(x,w)=1$. Knowledge soundness ensures that an accepted proof corresponds to a valid witness, while zero-knowledge prevents the verifier from learning~$w$ beyond what follows from the verified statement~\cite{groth2016size}. In \textsc{ShadowPath}, the holder acts as the prover and the credential verifier checks the proof. The private witness includes the credential's registry index and the authentication data required to establish its status under~$R_e$. The holder proves that the selected registry position is authenticated by~$R_e$ and represents a non-revoked credential. The verifier learns that the credential is not revoked without learning the registry index or authentication path.

\begin{figure*}[!t]
    \centering
    \definecolor{tmblue}{HTML}{2F5597}
\definecolor{tmgray}{HTML}{555B61}
\definecolor{tmborder}{HTML}{B8BDC3}
\definecolor{tmgreen}{HTML}{4E8B73}
\definecolor{tmgreenfill}{HTML}{F2F8F5}
\definecolor{tmorange}{HTML}{C98200}
\definecolor{tmorangefill}{HTML}{FFF8E8}
\definecolor{tmredfill}{HTML}{FBF5F5}

\begin{tikzpicture}[
  x=1cm,
  y=1cm,
  font=\footnotesize,
  actor/.style={
    draw=tmgray,
    rounded corners=2pt,
    line width=.7pt,
    minimum width=1.85cm,
    minimum height=.62cm,
    align=center,
    fill=white,
    font=\bfseries\small
  },
  card/.style={
    draw=tmborder,
    rounded corners=2.2pt,
    line width=.55pt,
    minimum width=3.25cm,
    minimum height=1.55cm,
    fill=white,
    inner sep=0pt
  },
  cardtitle/.style={
    font=\bfseries\footnotesize,
    align=center
  },
  cardbody/.style={
    font=\footnotesize,
    text=tmgray,
    align=center
  },
  protected/.style={
    font=\bfseries\footnotesize,
    text=tmgreen,
    align=center
  },
  residual/.style={
    font=\bfseries\footnotesize,
    text=tmorange,
    align=center
  },
  publicarrow/.style={
    -{Latex[length=1.6mm,width=1.05mm]},
    draw=tmblue,
    line width=.8pt
  },
  privatearrow/.style={
    -{Latex[length=1.6mm,width=1.05mm]},
    draw=tmgreen,
    line width=.75pt
  },
  rootarrow/.style={
    -{Latex[length=1.6mm,width=1.05mm]},
    draw=tmblue,
    densely dotted,
    line width=.8pt
  },
  mapping/.style={
    draw=tmgray!65,
    densely dotted,
    line width=.65pt
  },
  flowlabel/.style={
    fill=white,
    inner sep=1pt,
    align=center,
    font=\footnotesize
  }
]

\path[use as bounding box] (-0.4,0.12) rectangle (14.66,3.67);

\node[actor,fill=tmblue!6]
  (issuer) at (1.65,2.62) {Issuer};

\node[actor,fill=tmblue!4]
  (state) at (5.25,2.62) {Public state};

\node[actor,fill=tmgreenfill]
  (holder) at (8.9,2.62) {Holder};

\node[actor,fill=tmredfill]
  (verifier) at (12.44,2.62) {Verifier};

\draw[publicarrow]
  (issuer)--node[flowlabel, above=8pt, yshift=-0.1pt]
  {publish epoch state}(state);

\draw[publicarrow,densely dotted]
  (state)--node[flowlabel, above=6pt, yshift=2.2pt]
  {state sync}(holder);

\draw[privatearrow]
  (holder)--node[flowlabel, above=2pt, yshift=6pt]
  {bound status proof $\pi_{\mathsf{status}}$}(verifier);

\draw[rootarrow]
  (state.north)--++(0,.48)-|
  node[flowlabel, above, pos=0.62]
  {resolve~$(e,R_e)$}
  (verifier.north);

\node[card, minimum width=107pt, yscale=1.2] (issuerthreat) at (1.5,1.2) {};

\node[cardtitle] at (1.5,1.85)
  {Honest-but-curious issuer};

\node[cardbody] at (1.44,1.27)
  {Observes protocol traffic};

\node[protected, text width=112pt] at (1.59,0.6)
  {No credential-specific\\ query at presentation};

\draw[draw=tmgreen,line width=1pt,line cap=round]
  (-0.25,0.65)--(-0.17,0.56)--(-0.01,0.76);

\node[card, fill=tmorangefill, minimum width=102pt, yscale=1.2]
  (networkthreat) at (5.25,1.2) {};

\node[cardtitle] at (5.15,1.85)
  {Network observer};

\node[cardbody] at (5.27,1.3)
  {Observes retrieval metadata};

\node[residual, text width=100pt] at (5.3,0.6)
  {Residual IP/timing\\leakage};

\fill[tmorange]
  (3.67,0.6)--(3.78,0.79)--(3.89,0.6)--cycle;

\node[card, fill=tmgreenfill, minimum width=102pt, yscale=1.2]
  (holderthreat) at (8.9,1.2) {};

\node[cardtitle] at (8.86,1.85)
  {Malicious holder};

\node[cardbody, minimum width=87pt] at (8.89,1.25)
  {Attempts proofs for revoked\\credentials or forged paths};

\node[protected, text width=68pt, minimum width=0.9pt] at (9,0.61)
  {Revoked/forged proofs rejected};

\draw[draw=tmgreen,line width=1pt,line cap=round]
  (7.4,0.65)--(7.48,0.55)--(7.64,0.76);

\node[card, fill=tmgreenfill, minimum width=99pt, yscale=1.2, xscale=1.1]
  (verifierthreat) at (12.65,1.2) {};

\node[cardtitle] at (12.65,1.8)
  {Colluding verifiers};

\node[cardbody, minimum width=99pt] at (12.6,1.25)
  {Attempt cross-presentation\\linking};

\node[protected, text width=121pt] at (12.8,0.6)
  {Fresh session values;\\lookup remains hidden};

\draw[draw=tmgreen,line width=1pt,line cap=round]
  (10.9,0.65)--(10.98,0.56)--(11.14,0.77);

\draw[mapping] (1.5,2.31)--(issuerthreat.north);
\draw[mapping] (state.south)--(networkthreat.north);
\draw[mapping] (holder.south)--(holderthreat.north);
\draw[mapping] (verifier.south)--(12.44,2.08);

\end{tikzpicture}
    \vspace{-2.5mm}
    \caption{Threat-model overview. The issuer publishes authenticated epoch state, the holder synchronizes the shared state and generates a bound status proof, and the verifier independently resolves the selected epoch root. \textsc{ShadowPath} hides credential-specific lookup information from verifiers and avoids credential-specific issuer queries during presentation. State synchronization remains observable to network observers.}
    \label{fig:threat-model}
\end{figure*}
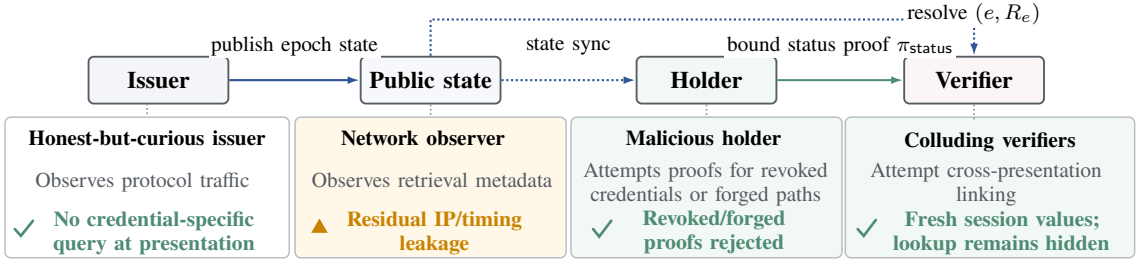

\section{Threat Model}\label{sec:threat-model}

\textsc{ShadowPath} considers the status-verification phase of a decentralized credential system with issuers, holders, verifiers, and a public state-distribution layer. Figure~\ref{fig:threat-model} shows the protocol flow, the adversaries considered, and the information available at their observation points.

\subsection{System and Adversary Model}
The issuer maintains the revocation registry as a sequence of immutable epochs~\cite{lueks2017fast}. For each epoch~$e$, the issuer publishes authenticated registry state with root~$R_e$. The holder synchronizes this shared state and derives the credential-specific authentication data locally. During presentation, the verifier independently resolves the selected epoch and root~$(e,R_e)$ according to its freshness policy. The holder generates a status proof against~$R_e$, and the verifier checks the proof without contacting an issuer-operated status responder~\cite{10.1145/3167132.3167303}.

Within an issuer domain, all credentials are evaluated under the same epoch share~$R_e$. The root identifies the registry state rather than an individual credential. Repeated credential identifiers, registry indices, traversal paths, authentication data, or retrieval metadata can become stable handles across presentations. \textsc{ShadowPath} hides the credential-specific lookup from the verifier, while the issuer domain and selected epoch remain public. State-synchronization metadata remains outside the cryptographic privacy guarantee.

We consider four computationally bounded adversaries~\cite{10.1007/978-3-031-30731-7_6}. An \emph{honest-but-curious issuer} publishes correct registry state but records visible protocol interactions and attempts to infer credential usage~\cite{DBLP:journals/popets/KrulPRK24}. \textsc{ShadowPath} removes the credential-specific issuer query during presentation. Synchronization traffic is considered separately at the network layer.

A \emph{passive network observer} monitors state synchronization and may correlate IP addresses, request timing and size, provider discovery, or retrieval behavior~\cite{10.1145/3544216.3544232,10.1145/3618257.3624797,10.1145/3672608.3707703}. A just-in-time synchronization may expose timing information. The observer cannot break encrypted transport or the assumed cryptographic primitives.

A \emph{malicious holder} controls the prover inputs and attempts to obtain an accepting proof for a revoked credential or a forged registry path~\cite{10.1145/3634737.3637641}. A sound status proof must reject both cases.

\emph{Colluding verifiers} retain and compare accepted presentations in an attempt to determine whether the same credential was used more than once. They observe the public presentation values and proof, but not the hidden credential identifier, registry index, traversal path, or authentication data.

\subsection{Security Assumptions and Scope}

Our guarantees rely on the following assumptions:

\begin{itemize}[leftmargin=*, itemsep=0.5mm, parsep=0pt]

\item \emph{Authenticated state and proof security.}
The issuer publishes correct roots and revocation updates. The authenticated-registry backend provides a binding root and sound path authentication. The ZKP system provides completeness, knowledge soundness, and zero-knowledge under sound setup parameters~\cite{10.1145/3335741.3335750,groth2016size,cryptoeprint:2019/953}. A valid composition binds the credential proof and status proof to the same hidden credential identifier and presentation session.

\item \emph{Fresh presentation values.}
Each presentation uses a fresh verifier challenge and fresh holder randomness to derive a new public session value. We assume independently generated session values are pseudorandom when the hidden credential identifier is unknown to the verifier. A verifier that learns a candidate identifier may test it against the public session value; issuer--verifier collusion is therefore excluded. Accepted challenges are consumed to prevent replay.

\item \emph{Observable state synchronization.}
State synchronization is credential-independent but not anonymous. Network observers may learn IP addresses, timing, routing information, and retrieval patterns. \textsc{ShadowPath} provides no private retrieval, cover traffic, or anonymous transport.

\end{itemize}

Malicious issuer behavior, omitted or delayed revocation updates, equivocation, availability, denial-of-service resistance, endpoint compromise, registry governance, and global state consistency are outside the threat model.

\subsection{Security and Privacy Objectives}

\textsc{ShadowPath} has one security objective, \emph{status correctness}, and three privacy objectives: \emph{lookup/path privacy}, \emph{issuer-query unobservability}, and \emph{presentation unlinkability}.

\noindent\textbf{Status correctness.}
An accepted status proof must establish that the presented credential is not revoked under the verifier-selected epoch root~$R_e$~\cite{lueks2017fast,10.1145/3167132.3167303}. The proof must also bind this status result to the credential presented in the same session. A malicious holder must be unable to obtain an accepting proof for a revoked credential or a forged registry path.

\noindent\textbf{\circRom{i} Lookup/path privacy.}
An accepted proof indicates that the credential is not revoked but does not reveal the credential-specific registry index, traversal path, or authentication data~\cite{11264637}. These values remain hidden inside the ZKP.

\noindent\textbf{\circRom{ii} Issuer-query unobservability.}
Status verification requires no credential-specific request to the issuer or an issuer-operated status service during presentation~\cite{10.1145/3319535.3354223}. State synchronization may still be visible, but it is shared across credentials rather than driven by a credential-specific lookup.

\noindent\textbf{\circRom{iii} Presentation unlinkability.}
With fresh pseudorandom session values, the verifier-visible status data do not reveal whether two accepted presentations use the same credential. Colluding verifiers receive no stable credential identifier, registry position, path, or authentication data. This guarantee excludes issuer--verifier collusion and does not cover state-synchronization metadata~\cite{9230363}.
\section{Problem Statement}\label{sec:problem}
\textsc{ShadowPath} addresses the following problem: a holder must prove that a credential is not revoked without revealing which registry entry was checked. The verifier must still establish non-revocation under the selected registry state and bind the result to the credential presented in the same session. Hiding the registry index or traversal path alone does not establish these bindings.

For each epoch~$e$, the issuer publishes an authenticated registry root~$R_e$. The holder uses the corresponding registry state to determine the credential's status locally and construct a proof. The proof must reveal neither the registry position, traversal path, nor authentication data. The status relation uses the same binding and non-revocation semantics across authenticated-registry backends. Path authentication depends on the selected backend.

Each proof uses a hidden credential identifier $VC_{\mathrm{id}}$. The holder derives it from the link secret $l_s$, issuer domain $I$, and issuer-authenticated nonce $\nu$. The same identifier determines the credential fingerprint $fp$ and private registry index $idx$.
Presentation binding adds a fresh verifier challenge $c$ and holder randomizer $r$ for the selected epoch $(e,R_e)$. The credential and status proofs must use the same hidden $VC_{\mathrm{id}}$ and enforce the public session-binding value
\noindent\makebox[\columnwidth][c]{\small
$\beta=H_{\mathsf{ent}}\!\left(
H_{\mathsf{bind}}\!\left(
VC_{\mathrm{id}},H_{\mathsf{ctx}}(c,e)\right),r\right).$
\quad
\textcolor{ieeeblue}{$\triangleright$}\;\textbf{public}}\par
Here, $H_{\mathsf{ctx}}$ binds the verifier challenge to the epoch, $H_{\mathsf{bind}}$ binds this context to the hidden credential identifier, and $H_{\mathsf{ent}}$ incorporates the holder randomizer. Fresh~$r$ produces a new~$\beta$ even if $(c,e)$ repeats. Binding to~$c$ prevents cross-challenge reuse, while consuming accepted challenges prevents direct replay. Presentation unlinkability relies on the session-value pseudorandomness assumption from \autoref{sec:threat-model} as collision resistance alone does not provide this property.

The protocol must satisfy three requirements. First, an accepted proof must establish non-revocation under the verifier-selected root and bind the result to the credential presented in the same session. Second, the protocol must provide the three privacy objectives: lookup/path privacy, issuer-query unobservability, and presentation unlinkability. Third, for fixed registry and circuit parameters, proof size and verification work must remain independent of the number of revoked credentials. 
For a formal statement, fix an authenticated-registry backend~$\mathcal D$ and issuer domain~$I$. Define:

\begingroup
\newcommand{\formulanote}[1]{%
  {\normalsize
   \textcolor{ieeeblue}{$\triangleright$}\;#1}%
}

\begin{center}
\renewcommand{\arraystretch}{1.25}
\small
\begin{tabular}{@{}c@{\hspace{0.35em}}l@{}}

$x=(R_e,e,c,r,\beta)$
&
\formulanote{\textbf{public} statement}
\\

$w=(ls,\nu,VC_{\mathrm{id}},\rho_{\mathcal D})$
&
\formulanote{\begin{tabular}[t]{@{}l@{}}complete \textbf{private}\\circuit witness\end{tabular}}
\\

$z=(\mathsf{fp},\mathsf{idx})$
&
\formulanote{\textbf{private}, derived}
\\[0.3ex]

$\displaystyle\begin{gathered}
 \mathsf{B}_{I}=
 \mathsf{Bind}_{I}(ls,\nu,VC_{\mathrm{id}},z;\\[-0.2ex]
 \hfill e,c,r,\beta)
\end{gathered}$
&
\formulanote{binding check}
\\

$\mathsf{A}_{\mathcal D}
 =\mathsf{AuthNonRevoked}_{\mathcal D}
   (R_e,\rho_{\mathcal D},z)$
&
\formulanote{status check}
\\[0.5ex]

$\displaystyle
 \begin{gathered}
   \mathcal{R}_{\mathcal D}(x,w)=1
      \iff \exists\,z:\vphantom{\mathsf{B}_{I}}\\[-0.1ex]
   \mathsf{B}_{I}=1
      \land
   \mathsf{A}_{\mathcal D}=1
 \end{gathered}$
&
\formulanote{status relation}

\end{tabular}
\end{center}
\endgroup

The binding predicate~$\mathsf{B}_{I}$ verifies that~$VC_{\mathrm{id}}$ is derived from $(ls,I,\nu)$, derives the private lookup tuple $z=(\mathsf{fp},\mathsf{idx})$, and checks the public session-binding value~$\beta$. The status predicate~$\mathsf{A}_{\mathcal D}$ authenticates the backend-specific witness~$\rho_{\mathcal D}$ under~$R_e$ and checks that the authenticated registry state does not revoke the credential identified by~$z$. The relation~$\mathcal{R}_{\mathcal D}$ accepts when both checks succeed. Credential composition additionally authenticates the issuer and requires the credential proof and status proof to use the same hidden~$VC_{\mathrm{id}}$. Table~\ref{tab:notation} summarizes the core notation used throughout the construction (see full notation in Appendix Table~\ref{tab:algorithm-environment}). 

\begin{table}[ht!]
\centering
\caption{Notation summary}
\label{tab:notation}
\footnotesize
\renewcommand{\arraystretch}{1.08}
\setlength{\tabcolsep}{3pt}
\begin{tabular}{@{}l p{0.76\columnwidth}@{}}
\toprule
\textbf{Symbol} & \textbf{Description} \\
\midrule
$e,\ R_e$ & Epoch; authenticated registry root \\
$c,\ r,\ \beta$ & Verifier challenge; holder randomizer; public session-binding value \\
$ls,\ \nu,\ VC_{\mathrm{id}}$ & Link secret; credential nonce; hidden credential identifier \\
$\mathsf{fp},\ \mathsf{idx}$ & Credential fingerprint; private registry index \\
$\mathcal D,\ \rho_{\mathcal D}$ & Registry backend; private authentication witness \\
$\mathsf{sum}_e,\ \Delta_e$ & Complete summary; authenticated delta for epoch~$e$ \\
$x,\ w,\ z$ & Public statement; complete private circuit witness; private, derived lookup tuple \\
$k,\ d$ & Registry branching factor; depth \\
$N_{\mathrm{rev}},\ b$ & Revoked-entry count; update-batch size \\
$\mathbb F,\mathbb F^\ast$
&
Outer circuit field $\mathbb F=\mathbb F_r(\mathrm{BW6\text{-}761})$; its nonzero elements
\\
\bottomrule
\end{tabular}
\end{table}

The registry path must be authenticated inside the proof to preserve both soundness and lookup privacy. Verifier-side path authentication would expose the path, while holder-side authentication outside the proof would require a trusted-wallet assumption and would not protect against a malicious holder. 

\begin{table*}[!t]
\caption{Comparison of revocation mechanisms}
\label{tab:rev_comparison}
\centering
\scriptsize
\setlength{\tabcolsep}{0.9pt}
\setlength{\shadowfillgap}{%
  \dimexpr(\textwidth-12.36cm-16\tabcolsep)/8\relax
}
\renewcommand{\arraystretch}{1.06}

\begin{NiceTabular*}{\textwidth}{
@{\extracolsep{\fill}}
>{\raggedright\arraybackslash}p{3.35cm}
>{\centering\arraybackslash}p{1.32cm}
>{\centering\arraybackslash}p{1.65cm}
>{\centering\arraybackslash}p{1.48cm}
>{\centering\arraybackslash}p{1.80cm}
>{\centering\arraybackslash}p{1.40cm}
>{\centering\arraybackslash}p{2.00cm}
>{\centering\arraybackslash}p{2.12cm}
>{\centering\arraybackslash}p{2.12cm}
@{}
}
\CodeBefore
  \rowcolor{gray!6}{28}
\Body

\toprule
\multicolumn{1}{c}{\textbf{Revocation Mechanism}}
&
\multicolumn{3}{c}{\textbf{Privacy Objectives}}
&
\multicolumn{3}{c}{\textbf{Operational Distinction}}
&
\multicolumn{2}{c}{\textbf{Revocation Data}}
\\
\cmidrule(lr){1-1}
\cmidrule(lr){2-4}
\cmidrule(lr){5-7}
\cmidrule(lr){8-9}
&
\begin{tabular}[t]{@{}c@{}}
\circRom{i}\\
Lookup/Path\\
Privacy
\end{tabular}
&
\begin{tabular}[t]{@{}c@{}}
\circRom{ii}\\
Issuer-Query\\
Unobservability
\end{tabular}
&
\begin{tabular}[t]{@{}c@{}}
\circRom{iii}\\
Presentation\\
Unlinkability\textsuperscript{b}
\end{tabular}
&
\begin{tabular}[t]{@{}c@{}}
Holder Update/\\
Access
\end{tabular}
&
\begin{tabular}[t]{@{}c@{}}
Shared Batch\\
Delta\textsuperscript{a}
\end{tabular}
&
\begin{tabular}[t]{@{}c@{}}
Root-Authenticated\\
Hidden Path
\end{tabular}
&
\begin{tabular}[t]{@{}c@{}}
With Prior\\
State
\end{tabular}
&
\begin{tabular}[t]{@{}c@{}}
Without Prior\\
State
\end{tabular}
\\
\midrule

\multicolumn{9}{l}{
\circNum{1}\ \textit{Status lists, compressed sets, and online responders}
}
\\[0.2ex]

\mbox{CRLs~\cite{10.1145/2815675.2815685}}
&
\ding{55} & \ding{51} & \ding{55}
&
\mbox{Shared sync}
&
\ding{55}
&
---
&
Full list
&
Full list
\\

\mbox{Bitstring Status List~\cite{W3C_Bitstring2024}}
&
\ding{55} & \ding{51} & \ding{55}
&
\mbox{Shared sync}
&
\ding{55}
&
---
&
Full list
&
Full list
\\

\mbox{Let's Revoke~\cite{smith2020let}}
&
\ding{55} & \ding{51} & \ding{55}
&
\mbox{Shared sync}
&
\ding{51}
&
---
&
CRV delta
&
Full CRV
\\

\mbox{CRLite (Bloom cascade)~\cite{7958597}}
&
\ding{55} & \ding{51} & \ding{55}
&
\mbox{Shared sync}
&
\ding{51}
&
---
&
\mbox{\qty{580}{\kilo\byte}/day}\textsuperscript{g}
&
\mbox{\qty{10}{\mega\byte}}\textsuperscript{g}
\\

\mbox{CRLite (clubcards)~\cite{schanck2025clubcards}}
&
\ding{55} & \ding{51} & \ding{55}
&
\mbox{Shared sync}
&
\ding{51}
&
---
&
\mbox{\qty{26.8}{\kilo\byte}/\qty{6}{\hour}}\textsuperscript{g}
&
\mbox{\qty{6.7}{\mega\byte}}\textsuperscript{g}
\\

\mbox{CTng~\cite{kong2026ctng}}
&
\ding{55} & \ding{51} & \ding{55}
&
\mbox{Shared sync}
&
\ding{51}
&
---
&
CRV delta
&
Full CRV
\\

\mbox{OCSP responders~\cite{10.1145/3278532.3278543}}
&
\ding{55} & \ding{55} & \ding{55}
&
\mbox{Online query}
&
\ding{55}
&
---
&
One reply
&
One reply
\\

\midrule
\multicolumn{9}{l}{
\circNum{2}\ \textit{Cryptographic accumulators and vector commitments}
}
\\[0.2ex]

\mbox{Dynamic Accumulators~\cite{camenisch2002dynamic}}
&
\ding{51}
&
\;(\ding{51})\textsuperscript{f}
&
\ding{51}
&
\mbox{Witness refresh}
&
\ding{55}
&
---
&
Witness update
&
One witness
\\

\mbox{Universal Accumulators~\cite{li2007universal}}
&
\ding{51}
&
\;(\ding{51})\textsuperscript{f}
&
\;(\ding{51})\textsuperscript{b}
&
\mbox{Witness refresh}
&
\ding{55}
&
---
&
Witness update
&
One witness
\\

\mbox{Pointproofs~\cite{10.1145/3372297.3417244}}
&
--- & --- & ---
&
\mbox{Proof update}
&
\ding{55}
&
---
&
---
&
---
\\

\mbox{ALLOSAUR~\cite{10.1145/3634737.3637641}}
&
\ding{51}
&
\;(\ding{51})\textsuperscript{f}
&
\ding{51}
&
\mbox{Oblivious refresh}
&
\ding{55}
&
---
&
Sublinear batch\textsuperscript{c}
&
One witness
\\

\mbox{AccuRevoke~\cite{ala_muid2025accurevoke}}
&
\ding{55}
&
\ding{51}\textsuperscript{d}
&
\ding{55}
&
\mbox{Edge query}
&
\ding{55}
&
---
&
One edge reply
&
One edge reply
\\

\mbox{EVOKE~\cite{mazzocca2024evoke}}
&
\ding{55}
&
\ding{51}
&
\ding{55}
&
\mbox{Witness refresh}
&
\ding{55}
&
---
&
Witness update
&
One witness
\\

\midrule
\multicolumn{9}{l}{
\circNum{3}\ \textit{Ledger-based revocation and identity systems}
}
\\[0.2ex]

\mbox{Sora Identity~\cite{8377927}}
&
\ding{55} & \ding{51} & \ding{55}
&
\mbox{Ledger access}
&
\ding{55}
&
---
&
One ledger record
&
One ledger record
\\

\mbox{CanDID~\cite{9519473}}
&
\ding{55} & \ding{51} & \ding{55}
&
\mbox{Public list}
&
\ding{55}
&
---
&
---
&
---
\\

\mbox{zkRevoke~\cite{manimaran2026zkRevoke}}
&
\ding{55}
&
\ding{51}
&
\ding{55}\textsuperscript{i}
&
\mbox{Offline holder}
&
\ding{55}
&
---
&
One status token\textsuperscript{i}
&
One status token\textsuperscript{i}
\\

\mbox{UPPR (AC mode)~\cite{11264637}}
&
\ding{51}
&
\ding{51}
&
\ding{51}
&
\mbox{Offline holder}
&
\ding{55}
&
---
&
Verifier-side\textsuperscript{e}
&
Verifier-side\textsuperscript{e}
\\

\midrule
\multicolumn{9}{l}{
\circNum{4}\ \textit{Zero-knowledge state trees and lookup arguments}
}
\\[0.2ex]

\mbox{V'CER~\cite{280040}}
&
\ding{55}
&
\ding{51}\textsuperscript{h}
&
\ding{55}
&
\mbox{Peer repair}
&
\ding{51}
&
\ding{55}
&
Changed paths
&
One tree path
\\

\mbox{SMT + ZKP (Iden3)~\cite{iden3}}
&
\ding{51}
&
\ding{51}
&
\;(\ding{51})\textsuperscript{b}
&
\mbox{Path refresh}
&
\ding{55}
&
\ding{51}
&
One tree path
&
One tree path
\\

\mbox{Merkle + ZKP (zk-creds)~\cite{zk-creds}}
&
\ding{51}
&
\ding{51}
&
\ding{51}
&
\mbox{Path refresh}
&
\ding{55}
&
\ding{51}
&
One tree path
&
One tree path
\\

\mbox{Caulk (Lookup Argument)~\cite{10.1145/3548606.3560646}}
&
\ding{51}
&
---
&
---
&
\mbox{Table access}
&
\ding{55}
&
---
&
---
&
---
\\

\mbox{\textbf{\textsc{ShadowPath} (Ours)}}
&
\textbf{\ding{51}}
&
\textbf{\ding{51}}
&
\textbf{(\ding{51})}\textsuperscript{b}
&
\mbox{\textbf{Shared sync}}
&
\textbf{\ding{51}}
&
\textbf{\ding{51}}
&
\textbf{Authenticated delta}
&
\textbf{Complete summary}
\\

\bottomrule
\end{NiceTabular*}

\begin{minipage}{0.99\textwidth}
\vspace{0.45mm}
\scriptsize
\raggedright
The revocation-data columns report what the holder or verifier must obtain or access. ``With prior state'' assumes retained local state; ``Without prior state'' denotes bootstrap or recovery. A dash denotes an inapplicable or undefined property. \textbf{\ding{51}} denotes that the property is fulfilled. \textbf{\ding{55}} denotes that the property is not fulfilled. 
\textsuperscript{a} One authenticated delta updates credential-independent shared holder state.
\textsuperscript{b} Presentation unlinkability requires fresh presentation randomness. 
\textsuperscript{c} ALLOSAUR transfers sublinear data relative to the update batch.
\textsuperscript{d} AccuRevoke avoids a CA query but exposes a credential-specific query to an ECP.
\textsuperscript{e} UPPR stores its Bloom-filter cascade at the verifier.
\textsuperscript{f} Privacy depends on the witness-distribution channel.
\textsuperscript{g} CRLite values are representative transfer sizes.
\textsuperscript{h} V'CER avoids per-validation CA queries; a node may contact the CA when collaborative proof repair fails.
\textsuperscript{i} zkRevoke requires no holder-side registry synchronization. Verifiers retrieve an epoch blocklist of $O(N_{\mathrm{rev}})$ tokens. Its core protocol permits repeated verification within a holder-selected period and does not provide presentation unlinkability.
\textit{Abbreviations:} AC, anonymous credential; CA, certificate authority; CRV, certificate revocation vector; ECP, edge compute provider; OCSP, Online Certificate Status Protocol.
\end{minipage}
\end{table*}

\section{Related Work}\label{sec:rel-work}
Credential-status mechanisms must keep revocation state up-to-date without turning queries, authentication paths, or proof transcripts into stable handles. Table~\ref{tab:rev_comparison} compares the three privacy objectives from \autoref{sec:threat-model}: \circRom{i} lookup/path privacy, \circRom{ii} issuer-query unobservability, and \circRom{iii} presentation unlinkability, together with update/access modes, operational properties, and state-transfer costs. Correctness and credential binding are baseline requirements and therefore omitted. Proof size is not normalized because distributed lists expose no holder-generated status proof, V'CER~\cite{280040} sends a logarithmic Merkle path, and zkRevoke's proof grows with the verification period~\cite{manimaran2026zkRevoke}. 


\noindent \circNum{1} \textbf{Status lists and online responders.} These mechanisms answer a credential-specific query or distribute status state before verification. OCSP~\cite{rfc6960} uses certificate-specific requests, allowing direct-query responders to observe the certificate and request timing~\cite{10196098,10.1145/3278532.3278543}. Certificate Revocation Lists~\cite{10.1145/2815675.2815685} distribute state instead. Bloom filters~\cite{10.1145/362686.362692,5751342}, Bitstring Status Lists~\cite{W3C_Bitstring2024}, and CRLite~\cite{7958597} compress that state. H\"olzl et al.~\cite{10.1145/3167132.3167303} adapt private revocation to mobile eIDs. Let's Revoke~\cite{smith2020let}, clubcards~\cite{schanck2025clubcards}, and CTng~\cite{kong2026ctng} refine indexing or update distribution. Related vehicular designs use list and Bloom-filter models~\cite{haas2009design,qi2020privacy}. Pre-presentation distribution avoids credential-specific issuer queries, but the relying party may still learn the checked index.

\noindent \circNum{2} \textbf{Cryptographic accumulators and vector commitments.} Accumulator schemes replace an explicit list with a digest and a holder witness. Dynamic accumulators~\cite{camenisch2002dynamic} update membership witnesses, while universal accumulators~\cite{li2007universal} also support non-membership. Ozdemir et al.~\cite{ozdemir2020scaling} optimize an RSA set accumulator for SNARK-based verifiable state updates as an alternative to Merkle trees; their relation is not a credential-status protocol. Privacy spans the presentation proof and the refresh channel. Proof-side designs either hide accumulator membership in anonymous-credential proofs~\cite{10.1007/978-3-642-24712-5_1} or update credentials without persistent presentation identifiers~\cite{10.1145/3319535.3354223}. Fraser and Schneider~\cite{fraser2025formal} formally analyze this pattern in Hyperledger AnonCreds while abstracting ledger access and witness refresh. Refresh-side work reduces or relocates witness-maintenance costs through oblivious sublinear updates~\cite{10.1145/3634737.3637641}, IoT or edge-assisted refresh~\cite{mazzocca2024evoke,ala_muid2025accurevoke}, or updatable vector openings~\cite{10.1145/3372297.3417244}. These interfaces separate state publication from privacy-preserving session evidence.

\noindent \circNum{3} \textbf{Ledger-based and public-state systems.} These systems publish status records, tokens, or roots so that the holder need not contact the issuer during presentation. Sora Identity~\cite{8377927} uses a blockchain ledger. CanDID~\cite{9519473} uses committee-maintained public state, and broader self-sovereign identity (SSI) registries follow related publication models~\cite{10891701,DBLP:journals/popets/KrulPRK24}. Public availability addresses state distribution but does not determine which evidence reaches the verifier. zkRevoke~\cite{manimaran2026zkRevoke} uses deterministic epoch tokens against a public list. Tokens for a common epoch link overlapping presentations, while its untraceability goal applies after the selected period. In UPPR's anonymous-credential mode~\cite{11264637}, the holder proves an epoch token and the verifier checks a public Bloom-filter cascade. Anonymity-revocation systems enable authorized parties to identify the holder of an anonymous credential and can make this tracing process auditable through privacy-preserving smart contracts~\cite{li2019audit}. Other verifier interfaces include private record certification~\cite{usenix2026_receipt}, modular keyed-verification credentials~\cite{usenix2026_sigmars}, and smart-contract-backed credentials~\cite{CHENG2024103735}. These interfaces separate public-state publication from private session evidence. 

\noindent \circNum{4} \textbf{Private lookup and zero-knowledge state trees.} Private information retrieval occupies a different point in the systems design space. Checklist~\cite{kogan2021private} privately queries a server-held blocklist through two non-colluding servers. This avoids full-list storage and holder-side proving but requires online server computation. It does not produce a verifier-portable root-linked status proof. Zero-knowledge state trees instead authenticate private state against a public root. V'CER~\cite{280040} sends its sparse-Merkle path to the verifier or an assisting peer. Iden3~\cite{iden3, iden3protocol} and zk-creds~\cite{zk-creds} authenticate paths inside zero-knowledge. Iden3 hides a sparse-Merkle path, whereas zk-creds proves membership in a Merkle forest. ZEBRA~\cite{rathee2022zebra} combines sparse-Merkle non-membership with wallet authorization and accountable tracing. \textsc{ShadowPath} does not claim to provide the first private non-membership proof or to be the first system to hide a tree path. Its contributions are the credential-independent epoch-state lifecycle, the verifier-selected root interface, the same-credential/session composition, and the measured lifecycle costs. ZEBRA does not define private-witness retrieval or refresh, so we use it as a proof-side comparison without inferring transfer costs. Verkle trees~\cite{kuszmaul2019verkle} use vector commitments~\cite{catalano2013vector} to reduce authentication-path length, while Caulk~\cite{10.1145/3548606.3560646} and Lasso~\cite{setty2024unlocking} provide efficient lookup arguments that can be integrated into larger protocols. These constructions can hide the lookup index and authentication path within a circuit. End-to-end privacy still depends on how holders obtain and update their witnesses.

\noindent\textbf{Domain-specific and adjacent revocation.} Domain-specific work changes the revocation object or communication model, so its privacy goals do not map directly to Table~\ref{tab:rev_comparison}. Sitouah et al.~\cite{10634387} rerandomize affected Merkle subsets while presenting a credential identifier and authentication data. Their notion of untraceability differs from lookup/path privacy. Mobility-oriented systems prioritize dissemination and timely enforcement. V2X work covers pseudonym provisioning~\cite{brecht2018security}, distributed revocation~\cite{naskar2024scheme}, trusted-component protocols~\cite{whitefield2017formal}, and bounded-time self-revocation~\cite{scopelliti2024efficient}. Mobile-gateway IoT work addresses revocation of cooperation across a gateway topology~\cite{zhou2022perils}. Other domains revoke TLS certificates~\cite{10.1145/3785653}, group-signing rights~\cite{chu2012verifier}, DNS delegations~\cite{li2023ghost}, or offline double-spenders~\cite{dmitrienko2017secure}. 

\noindent\textbf{Summary.} The systems closest to \textsc{ShadowPath} fall into two groups. Iden3~\cite{iden3}, zk-creds~\cite{zk-creds}, and ZEBRA~\cite{rathee2022zebra} are closest to its root-authenticated hidden lookup. UPPR~\cite{11264637} and zkRevoke~\cite{manimaran2026zkRevoke} are closest to its presentation-independent distribution of status evidence. The systems problem addressed here is to preserve one privacy argument across credential-independent state acquisition, root-linked non-revocation, and binding to the current credential presentation.

\textsc{ShadowPath} addresses this task via a single protocol interface. Holders synchronize credential-independent epoch state. The circuit authenticates a hidden lookup under the verifier-selected root, and session binding ties the status proof to the current presentation. Together, these steps avoid credential-specific traffic to the issuer during presentation without exposing the registry position. Synchronization metadata remains outside the cryptographic guarantee. The contribution is this integrated workflow and its evaluated trade-offs, not a new cryptographic primitive. Architecturally, \textsc{ShadowPath} is closest to the zero-knowledge state-tree group, while its credential-independent synchronization follows the shared-state distribution model used by related systems.
\section{\textsc{ShadowPath} Protocol and Registry Backends} \label{sec:implementation}

\subsection{Protocol Execution and Composition}
\textsc{ShadowPath} is independent of the authenticated-registry backend. SMT and Verkle use the same protocol flow and differ only in witness construction and path authentication. The evaluated Verkle implementation provides the complete prototype, including state publication, holder synchronization, local witness reconstruction, and proof generation. The SMT implementation is restricted to the controlled in-memory proof comparison used to isolate path-authentication costs. The Verkle prototype distributes authenticated epoch state through the InterPlanetary File System (IPFS) and InterPlanetary Name System (IPNS)~\cite{10.1145/3544216.3544232}. IPFS stores content-addressed epoch objects, while IPNS provides an authenticated pointer to the current epoch. Algorithm~\ref{alg:fetch} therefore presents the implemented Verkle state-retrieval and witness-reconstruction workflow.

\begin{algorithm}[!ht]
\caption{\textsc{ShadowPath}-Verkle State Retrieval and Local Witness Reconstruction}
\label{alg:fetch}
\footnotesize
\begin{algorithmic}[1]

\Statex \textbf{Actor:} Holder wallet
\Require Pinned issuer IPNS name $P_{\mathsf{ipns}}$,
private registry index $\mathsf{idx}$,
holder freshness state $\mathsf{st}_{\mathsf H}$,
freshness policy $\mathcal F$
\Ensure $(e,R_e,\rho_{\mathsf{Verkle}})$ or $\bot$

\State $\mathsf{cid}_e
\gets \mathsf{IPNS.ResolveVerified}(P_{\mathsf{ipns}})$
\Comment{Authenticate issuer \Statex \hfill epoch pointer}
\State $\mathsf{obj}_e
\gets \mathsf{IPFS.Fetch}(\mathsf{cid}_e)$
\Comment{Retrieve epoch object}

\If{$\mathsf{VerifyCID}(\mathsf{cid}_e,\mathsf{obj}_e)=0$}
    \State \Return $\bot$
\EndIf

\State $(\mathsf{payload}_e,\mathsf{checksum}_e)
\gets \mathsf{Deserialize}(\mathsf{obj}_e)$

\If{$\mathsf{Checksum}(\mathsf{payload}_e)
\neq \mathsf{checksum}_e$}
    \State \Return $\bot$
\EndIf

\State $(e,t_e,\mathsf{prev}_e,\mathsf{sum}_e)
\gets \mathsf{payload}_e$
\Comment{Extract epoch metadata and \Statex \hfill shared state}

\State $(R_e,\mathsf{state}_e)
\gets \mathsf{ParseSummary}(\mathsf{sum}_e)$

\If{$\mathcal F(\mathsf{st}_{\mathsf H},
e,t_e,\mathsf{prev}_e,R_e)=0$}
    \State \Return $\bot$
\EndIf

\State $\rho_{\mathsf{Verkle}}
\gets \mathsf{ReconstructWitness}_{\mathsf{Verkle}}
(\mathsf{idx},\mathsf{state}_e)$
\Comment{Derive private \Statex \hfill authentication witness}

\State $\mathsf{StoreFreshnessState}
(\mathsf{st}_{\mathsf H},e,R_e)$
\Comment{Record accepted epoch}

\State \Return $(e,R_e,\rho_{\mathsf{Verkle}})$

\end{algorithmic}
\end{algorithm}

Algorithm~\ref{alg:prove} describes proof generation from synchronized state after the verifier has selected~$(e,R_e)$ and issued~$c$. The holder returns $(\pi_{\mathsf{status}},r,\beta)$ for verification. The verifier runs neither algorithm. The verifier resolves the current root independently and verifies the proof under~$R_e$.

\begin{algorithm}[!t]
\caption{\textsc{ShadowPath} Non-Revocation Proof Generation}
\label{alg:prove}
\footnotesize
\begin{algorithmic}[1]

\Statex \textbf{Actor:} Holder wallet
\Require Registry backend $\mathcal D$,
link secret $ls$,
issuer domain $I$,
credential nonce $\nu$,
verifier-selected epoch $(e,R_e)$,
verifier challenge $c$,
authentication witness $\rho_{\mathcal D}$,
proving key $\mathsf{pk}_{\mathcal P}$,
proof backend $\mathcal P$
\Ensure $(\pi_{\mathsf{status}},r,\beta)$ or $\bot$

\State $VC_{\mathrm{id}}
\gets
H_{\mathsf{nonce}}\!\left(
H_{\mathsf{cred}}(ls,I),\nu
\right)$
\Comment{Hidden credential identifier}

\State $(\mathsf{fp},\mathsf{idx})
\gets
\mathsf{Split}\!\left(
H_{\mathsf{addr}}(VC_{\mathrm{id}})
\right)$
\Comment{Private lookup tuple}

\State $r\xleftarrow{\$}\mathbb F^\ast$
\Comment{Fresh holder randomness}

\State $\beta
\gets
H_{\mathsf{ent}}\!\left(
H_{\mathsf{bind}}\!\left(
VC_{\mathrm{id}},H_{\mathsf{ctx}}(c,e)
\right),r
\right)$
\Comment{Public session binding}

\State $x
\gets
(R_e,e,c,r,\beta)$

\State $w
\gets
(ls,\nu,VC_{\mathrm{id}},\rho_{\mathcal D})$

\State $\pi_{\mathsf{status}}
\gets
\mathcal P.\mathsf{Prove}\!\left(
\mathsf{pk}_{\mathcal P},
\mathcal R_{\mathcal D},
x,w
\right)$
\Comment{Generate status proof}

\If{$\pi_{\mathsf{status}}=\bot$}
    \State \Return $\bot$
\EndIf

\State \Return $(\pi_{\mathsf{status}},r,\beta)$
\Comment{Return proof and session values}

\end{algorithmic}
\end{algorithm}

Writing~$\mathsf V$ and~$\mathsf H$ for the verifier and holder, the logical protocol statement is
\begingroup
\newcommand{\formulanote}[1]{%
  {\normalsize
   \textcolor{ieeeblue}{$\triangleright$}\;#1}%
}

\begin{center}
\renewcommand{\arraystretch}{1.2}
\small
\begin{tabular}{@{}c@{\hspace{0.6em}}l@{}}

$\mathsf V\xrightarrow{(e,R_e,c)}\mathsf H$
&
\formulanote{verifier-selected context}
\\[0.4ex]

$\displaystyle
\begin{gathered}
\beta=H_{\mathsf{ent}}\!\bigl(
H_{\mathsf{bind}}\!
  (VC_{\mathrm{id}},H_{\mathsf{ctx}}(c,e)),r
\bigr).
\end{gathered}$
&
\formulanote{\textbf{public} session binding}

\end{tabular}
\end{center}
\endgroup
Acceptance requires the credential and status relations to enforce this derivation and use the same hidden $VC_{\mathrm{id}}$.

During issuance, the holder derives $VC_{\mathrm{id}}=H_{\mathsf{nonce}}(H_{\mathsf{cred}}(ls,I),\nu)$ and sends it to the issuer. Within a well-formed registry, the credential is revoked when the fingerprint stored at its registry position equals~$\mathsf{fp}$. An empty position or a different fingerprint establishes non-revocation. Issuance rejects $\mathsf{fp}=0$, reserves~$\mathsf{idx}$, and retries with a fresh~$\nu$ if the index is occupied. Index reservation ensures uniqueness, while the fingerprint check provides defense-in-depth. We scope this profile to at most $10^6$ reserved indices per issuer domain. Appendix Figure~\ref{fig:sp_bit_allocation} records the bit allocation, and Appendix~\ref{app:mode-b-derivation} analyzes collisions and wider profiles. Because the issuer learns~$VC_{\mathrm{id}}$, presentation unlinkability does not hold under issuer--verifier collusion.

Both backends establish non-revocation from an authenticated absent entry or an authenticated fingerprint different from~$\mathsf{fp}$. The SMT circuit recomputes the root from a hidden depth-50 path. The Verkle circuit verifies five KZG openings and links the selected termination level to~$R_e$. Both hide the registry index, traversal path, opened values, and backend-specific authentication data. The Verkle implementation verifies BLS12-377 pairings inside BW6-761 Groth16 and PLONK circuits. The BLS12-377 base field equals the BW6-761 scalar field, avoiding generic foreign-field emulation. Appendix Figure~\ref{fig:shadowpath_relation} illustrates the relation. See Appendices~\ref{app:mode-b-derivation} and~\ref{app:mode-b-implementation} for equations, point validation, and implementation details.

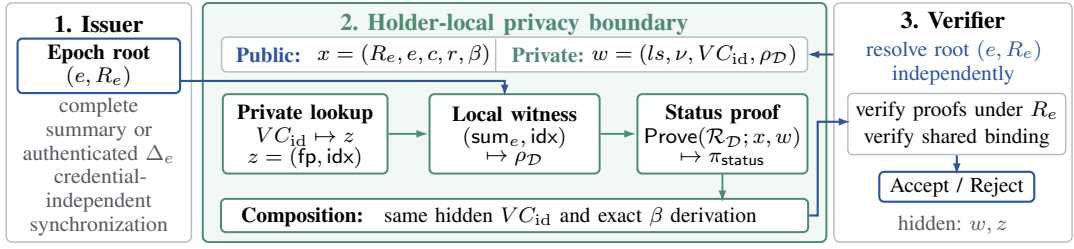
\begin{figure*}[!t]
    \centering
    \definecolor{archblue}{HTML}{2F5597}
\definecolor{archgray}{HTML}{555B61}
\definecolor{archborder}{HTML}{B8BDC3}
\definecolor{archgreen}{HTML}{4E8B73}
\definecolor{archgreenfill}{HTML}{F2F8F5}

\begin{tikzpicture}[
  x=1cm,
  y=1cm,
  font=\footnotesize,
  region/.style={
    draw=archborder,
    rounded corners=2.2pt,
    line width=.55pt,
    fill=white,
    inner sep=0pt
  },
  title/.style={
    font=\bfseries\small,
    align=center
  },
  box/.style={
    draw=archborder,
    rounded corners=1.8pt,
    line width=.55pt,
    fill=white,
    align=center,
    inner xsep=3pt,
    inner ysep=2pt
  },
  publicbox/.style={
    box,
    draw=archblue,
    line width=.8pt
  },
  privatebox/.style={
    box,
    draw=archgreen,
    line width=.75pt
  },
  btitle/.style={
    font=\bfseries\footnotesize,
    align=center
  },
  bline/.style={
    font=\footnotesize,
    align=center
  },
  publicarrow/.style={
    -{Latex[length=1.6mm,width=1.05mm]},
    draw=archblue,
    line width=.8pt
  },
  privatearrow/.style={
    -{Latex[length=1.6mm,width=1.05mm]},
    draw=archgreen,
    line width=.7pt
  },
  note/.style={
    text=archgray,
    align=center,
    font=\footnotesize
  }
]

\path[use as bounding box] (-0.03,0.21) rectangle (14.15,3.59);

\node[region, minimum width=69.71pt, minimum height=90pt] (state) at (1.25,1.90) {};

\node[title] at (1.25,3.2)
  {1. Issuer};

\node[
  publicbox,
  minimum width=2.15cm,
  minimum height=.72cm
] (epochstate) at (1.25,2.65) {};

\node[btitle] at (1.25,2.8)
  {Epoch root};

\node[bline] at (1.25,2.5)
  {$(e,R_e)$};

\node[note,text width=2.30cm] at (1.25,1.8)
  {complete summary or authenticated~$\Delta_e$};

\node[note,text width=2.30cm] at (1.25,0.85)
  {credential-independent\\synchronization};

\node[region, draw=archgreen, line width=.75pt, fill=archgreenfill, minimum width=234.74pt, minimum height=90pt] (holder) at (6.75,1.9) {};

\node[title,text=archgreen] at (6.75,3.2)
  {2. Holder-local privacy boundary};

\node[
  box,
  minimum width=7.75cm,
  minimum height=.46cm
] (header) at (6.75,2.8) {};

\draw[draw=archborder,line width=.5pt]
  (6.5,2.57)--(6.5,3.01);

\node[font=\bfseries\footnotesize,text=archblue,anchor=east]
  at (4.05,2.8) {Public:};

\node[anchor=west] at (4.02,2.8)
  {$x=(R_e,e,c,r,\beta)$};

\node[font=\bfseries\footnotesize,text=archgreen,anchor=east]
  at (7.8,2.8) {Private:};

\node[anchor=west] at (7.65,2.8)
  {$w=(ls,\nu,VC_{\mathrm{id}},\rho_{\mathcal D})$};

\node[privatebox, minimum width=62pt, minimum height=1.05cm] (lookup) at (3.98,1.75) {};

\node[btitle] at (3.98,2.02)
  {Private lookup};

\node[bline] at (3.98,1.71)
  {$VC_{\mathrm{id}}\mapsto z$};

\node[bline] at (3.98,1.43)
  {$z=(\mathsf{fp},\mathsf{idx})$};

\node[
  privatebox,
  minimum width=2.25cm,
  minimum height=1.05cm
] (witness) at (6.75,1.75) {};

\node[btitle] at (6.75,2.02)
  {Local witness};

\node[bline] at (6.75,1.7)
  {$(\mathsf{sum}_e,\mathsf{idx})$};

\node[bline] at (6.75,1.42)
  {$\mapsto\rho_{\mathcal D}$};

\node[privatebox, minimum width=64pt, minimum height=1.05cm] (proof) at (9.5,1.75) {};

\node[btitle] at (9.5,2.02)
  {Status proof};

\node[bline] at (9.5,1.7)
  {$\mathsf{Prove}(\mathcal R_{\mathcal D};x,w)$};

\node[bline] at (9.5,1.42)
  {$\mapsto\pi_{\mathsf{status}}$};

\draw[privatearrow] (lookup)--(witness);
\draw[privatearrow] (witness)--(proof);

\draw[publicarrow]
  (2.33,2.45)--(2.42,2.45)--
  (2.42,2.45)--(6.62,2.45)--(6.62,2.24);

\draw[publicarrow]
  (11,2.8)--(10.62,2.8);

\node[privatebox, minimum width=221pt, minimum height=.44cm] (composition) at (6.75,0.67) {};

\node[btitle,anchor=west] at (3.01,0.65)
  {Composition:};

\node[anchor=west, yshift=-0.3pt] at (4.9,0.67)
  {same hidden~$VC_{\mathrm{id}}$ and exact~$\beta$ derivation};

\draw[privatearrow]
  (proof.south)--(9.5,0.84);

\node[region, minimum width=89.63pt, minimum height=90pt] (verifier) at (12.55,1.90) {};

\node[title] at (12.55,3.25)
  {3. Verifier};

\node[note,text=archblue,text width=2.75cm] at (12.55,2.7)
  {resolve root $(e,R_e)$\\independently};

\node[box, minimum width=82pt, minimum height=23pt] (checks) at (12.6,1.88) {};

\node[bline] at (12.6,2.05)
  {verify proofs under~$R_e$};

\node[bline] at (12.6,1.69)
  {verify shared binding};

\node[
  publicbox,
  minimum width=2.05cm,
  minimum height=.34cm
] (decision) at (12.6,1.05)
  {Accept / Reject};

\draw[publicarrow]
  (checks)--(12.6,1.2);

\node[note] at (12.55,0.55)
  {hidden: $w,z$};

\draw[publicarrow]
  (composition.east)--(10.76,0.67)--
  (10.76,1.91)--(11.2,1.91);

\end{tikzpicture}
    \caption{\textsc{ShadowPath} holder-local architecture. Blue marks authenticated epoch state and public values. Pale green marks the hidden lookup and backend-specific witness~$\rho_{\mathcal D}$. The credential-authentication and status relations use the same hidden~$VC_{\mathrm{id}}$ and exact~$\beta$ derivation. The verifier receives the proofs and declared public values; $w$ and $z$ remain hidden.}
    \label{fig:shadowpath_overview}
\end{figure*}

Figure~\ref{fig:shadowpath_overview} shows the composition interface. The \textsc{ShadowPath}-Verkle proof of concept verifies a private issuer signature~$\sigma_I$ on~$VC_{\mathrm{id}}$ together with the complete status relation. Both checks use the same hidden identifier and enforce the exact~$\beta$ derivation. This artifact is not a production anonymous-credential integration. Such an integration must preserve both conditions within one sound composition mechanism. The verifier consumes $c$ after acceptance. 

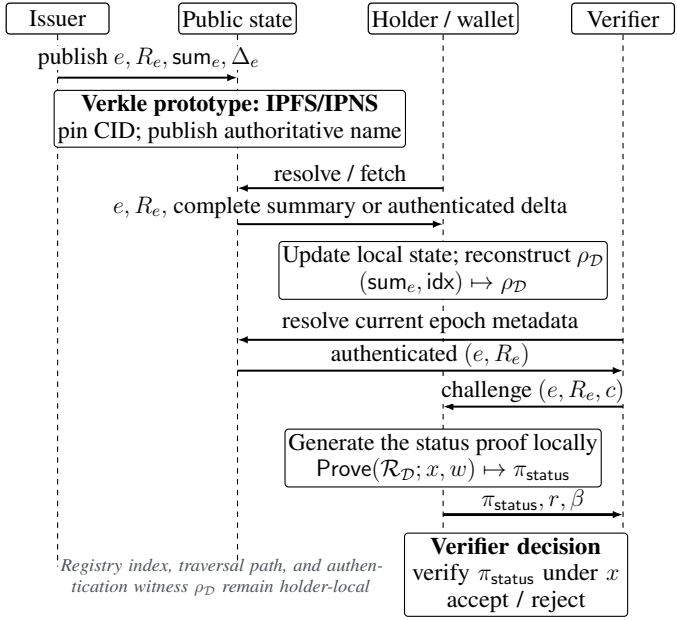
\begin{figure}[!ht]
    \centering
    \definecolor{spgray}{HTML}{4D5359}

\begingroup
\resizebox{\columnwidth}{!}{%
\begin{tikzpicture}[
  font={\rmfamily\fontsize{13.2pt}{14.6pt}\selectfont},
  actor/.style={
    draw=black,
    fill=white,
    rounded corners=2pt,
    line width=.7pt,
    minimum width=20mm,
    minimum height=6mm,
    align=center,
    inner xsep=2pt,
    inner ysep=2.5pt
  },
  process/.style={
    draw=black,
    fill=white,
    rounded corners=2pt,
    line width=.6pt,
    align=center,
    inner xsep=2pt,
    inner ysep=2.5pt
  },
  msg/.style={
    -{Latex[length=2mm,width=1.5mm]},
    draw=black,
    line width=1.1pt
  },
  proofmsg/.style={
    msg,
    line width=1.6pt
  },
  textlabel/.style={
    align=center,
    fill=white,
    inner xsep=1.5pt,
    inner ysep=.5pt
  },
  privacy/.style={
    text=spgray,
    align=center,
    font=\itshape
  }
]

\node[actor] (issuer) at (0.5,8.95) {Issuer};
\node[actor] (state) at (4,8.95) {Public state};
\node[actor] (holder) at (8,8.95) {Holder / wallet};
\node[actor] (verifier) at (11.5,8.95) {Verifier};

\foreach \n in {issuer,state,holder,verifier}
  \draw[black,dashed,line width=.6pt]
    (\n.south)--++(0,-10.35);

\draw[msg] (0.5,7.8) -- (4,7.8);
\node[textlabel,anchor=south] at (2.27,7.95)
  {publish $e,R_e,\mathsf{sum}_e,\Delta_e$};

\node[process,minimum width=32mm] at (3.85,7)
  {\textbf{Verkle prototype: IPFS/IPNS}\\
   pin CID; publish authoritative name};

\draw[msg] (8,5.65) -- (4,5.65);
\node[textlabel,anchor=south] at (6,5.75)
  {resolve / fetch};

\draw[msg] (4,4.95) -- (8,4.95);
\node[textlabel,anchor=south] at (6,5.05)
  {$e,R_e,$ complete summary or authenticated delta};

\node[process,minimum width=36mm] (reconstruct) at (8,4.05)
  {Update local state; reconstruct $\rho_{\mathcal D}$\\
   $(\mathsf{sum}_e,\mathsf{idx})
   \mapsto\rho_{\mathcal D}$};

\draw[msg] (11.5,2.7) -- (4,2.7);
\node[textlabel,anchor=south] at (7.75,2.8)
  {resolve current epoch metadata};

\draw[msg] (4,2.1) -- (11.5,2.1);
\node[textlabel,anchor=south] at (7.75,2.15)
  {authenticated $(e,R_e)$};

\draw[msg] (11.5,1.4) -- (8,1.4);
\node[textlabel,anchor=south] at (9.75,1.45)
  {challenge $(e,R_e,c)$};

\node[process,minimum width=36mm] (prove) at (8,0.4)
  {Generate the status proof locally\\
   $\mathsf{Prove}(\mathcal R_{\mathcal D};x,w)
   \mapsto\pi_{\mathsf{status}}$};

\draw[proofmsg] (8,-0.7) -- (11.5,-0.7);
\node[textlabel,anchor=south] at (9.75,-0.65)
  {$\pi_{\mathsf{status}},r,\beta$};

\node[
  process,
  text width=42mm,
  anchor=north east
] at (11.6,-1.05)
  {\textbf{Verifier decision}\\
   verify $\pi_{\mathsf{status}}$ under $x$\\
   accept / reject};

\node[
  privacy,
  text width=64mm,
  anchor=east
] at (7,-1.92)
  {Registry index, traversal path, and
   authentication witness~$\rho_{\mathcal D}$
   remain holder-local};

\end{tikzpicture}%
}
\endgroup
    \caption{\textsc{ShadowPath} protocol sequence. The Verkle prototype distributes state through the IPFS and IPNS. Holder-local boxes show state update, witness reconstruction, and proof generation. Here, CID denotes a content identifier.}
    \label{fig:sp_architecture}
\end{figure}

\textsc{ShadowPath} converts a credential-specific lookup over authenticated epoch state into a session-bound zero-knowledge proof of non-revocation. Figure~\ref{fig:sp_architecture} shows the protocol sequence. Before presentation, holders synchronize shared state and reconstruct their authentication witnesses locally. During presentation, they prove under the epoch root selected by the verifier.

\subsection{Relation and Registry Backends}

The relation~$\mathcal R_{\mathcal D}$ from \autoref{sec:problem} combines credential-identifier derivation and session binding with backend-specific authenticated non-revocation. Backend dependence is confined to $\mathsf{AuthNonRevoked}_{\mathcal D}$. It authenticates~$\rho_{\mathcal D}$ under~$R_e$ and establishes non-revocation from authenticated absence or fingerprint mismatch. The backend also determines serialization, setup, synchronization, and state distribution.

Table~\ref{tab:registry-backends} compares the two backends. \textsc{ShadowPath}-SMT recomputes a depth-50 Poseidon~\cite{poseidon} path from 50 sibling values. \textsc{ShadowPath}-Verkle uses a sparse KZG-authenticated tree over BLS12-377~\cite{kuszmaul2019verkle,kate2010constant}. With $k=1024$ and $d=5$, it covers the same $2^{50}$ positions with five openings. Both share the lookup derivation, session binding, fingerprint check, and absent-entry semantics. The comparison asks whether five KZG openings cost less than 50 hash levels.

\begin{table}[!t]
\centering
\caption{Authenticated-registry backends evaluated in \textsc{ShadowPath}.}
\label{tab:registry-backends}

\scriptsize
\setlength{\tabcolsep}{2.2pt}
\renewcommand{\arraystretch}{1.08}

\begin{tabularx}{\columnwidth}{@{}
  >{\raggedright\arraybackslash}p{0.22\columnwidth}
  >{\raggedright\arraybackslash}X
  >{\raggedright\arraybackslash}X@{}}
\toprule
\textbf{Property}
&
\textbf{\textsc{ShadowPath}-SMT}
&
\textbf{\textsc{ShadowPath}-Verkle}
\\
\midrule

Path commitment
&
Poseidon hash nodes
&
KZG vector commitments
\\

Geometry
&
Binary, $d=50$
&
$k=1024$, $d=5$
\\

Outer curve / field
&
BW6-761 / $\mathbb F_r$
&
BW6-761 / $\mathbb F_r$
\\

Private path
&
50 sibling values
&
5 node openings
\\

In-circuit check
&
Recompute hash root
&
Verify and link KZG openings
\\

Tree-specific setup
&
None
&
KZG setup\textsuperscript{b}
\\

Proof backend
&
Groth16\textsuperscript{a}
&
Groth16 and PLONK
\\

Artifact scope
&
Standalone relation; in-memory state control
&
Full service, composition proof of concept, and mobile prototype
\\

\bottomrule
\end{tabularx}

\begin{minipage}{0.99\linewidth}
\vspace{0.45mm}
\scriptsize
\raggedright
\textsuperscript{a} The matched SMT--Verkle comparison uses Groth16 for both backends; the PLONK evaluation covers Verkle.
\textsuperscript{b} Verkle uses BLS12-377 for KZG; its base field equals
the BW6-761 scalar field used by the outer proof circuit.
\end{minipage}

\end{table}

\subsection{Security Arguments}
We first consider the standalone status proof. Let~$\mathcal D$ denote the selected authenticated-registry backend, either SMT or Verkle. We assume that~$R_e$ is authenticated, policy-fresh, and derived from a well-formed registry under the honest-issuer lifecycle. The shared Poseidon derivations must be collision-resistant. The selected registry backend must provide complete and sound path authentication, while the outer proof system must provide completeness, knowledge soundness, and zero-knowledge under sound setup parameters.

For the SMT backend, path authentication relies on collision-resistant Poseidon hashing and correct root recomputation. For the Verkle backend, it additionally relies on KZG binding and evaluation soundness, collision resistance of the commitment encoding~$\mathsf{Enc}$, a sound KZG setup, and valid in-circuit curve and subgroup checks for all prover-supplied KZG points.

Under these assumptions, an honestly generated proof for a non-revoked credential is accepted. Knowledge soundness ensures that any accepted proof corresponds to a valid $(VC_{\mathrm{id}},\mathsf{fp},\mathsf{idx},\rho_{\mathcal D})$ that authenticates non-revocation under~$R_e$. Zero-knowledge hides the private registry lookup and authentication data in~$w$, including~$\rho_{\mathcal D}$, beyond what follows from the public statement~$x$. These guarantees are relative to the authenticated root~$R_e$ and do not imply registry completeness, timeliness, or global consistency.

\noindent\textit{Argument.}
An honest witness satisfies both predicates in $\mathcal{R}_{\mathcal D}$. Knowledge extraction yields a witness whose lookup tuple, authenticated path, and terminal condition satisfy the relation. Poseidon hashing binds an SMT path to~$R_e$, while KZG soundness and commitment linking provide the Verkle binding. Zero-knowledge hides the private witness in each accepted proof, while the session-value pseudorandomness assumption prevents independently generated session values from providing a stable credential handle. Together, these properties support presentation unlinkability as defined in \autoref{sec:threat-model}. This guarantee excludes issuer--verifier collusion and synchronization metadata.

\noindent\textbf{Credential-composition argument.}
Assume the standalone status conditions, an authenticated issuer key, and a sound credential-authentication predicate. A knowledge-sound, zero-knowledge composition proves that the issuer authenticated the hidden~$VC_{\mathrm{id}}$ and that the same~$VC_{\mathrm{id}}$ satisfies~$\mathcal R_{\mathcal D}$ under~$R_e$. Both checks enforce the exact session-binding derivation for~$\beta$. Acceptance therefore binds non-revocation under~$R_e$ to the credential authenticated in that presentation. Zero-knowledge hides~$VC_{\mathrm{id}}$ and the private registry witness beyond the public statement.

\emph{Argument.} Knowledge soundness extracts both witnesses. Identifier equality and the exact~$\beta$ derivation bind them to the same credential, epoch, and challenge. Binding to~$c$ prevents cross-challenge use. Consuming accepted challenges prevents replay, whereas joint zero-knowledge hides both witnesses.

\subsection{Artifact and Distribution Boundaries} The \textsc{ShadowPath}-Verkle prototype combines a Rust issuer and registry with Go/gnark services. Its Groth16 and PLONK circuits verify all KZG openings internally, and negative tests exercise point validation. The controlled SMT artifact is limited to the standalone Groth16 relation with in-memory state. It excludes composition, IPFS/IPNS distribution, and the mobile prototype. The two columns of Table~\ref{tab:registry-backends} are not equivalent deployments. Appendix~\ref{app:setup_retrieval} specifies setup and the Rust--Go encoding bridge.

The Verkle prototype publishes complete summaries and authenticated deltas as content-addressed IPFS objects. The pinned issuer IPNS name identifies the current epoch object. Kubo verifies the signed IPNS record and cryptographically binds each content identifier to its referenced object. A delta is not signed separately; publisher authentication follows from the signed epoch head and CID. Delta acceptance additionally requires the predecessor metadata and recomputed root to match the authenticated epoch object. The holder rejects stale or rolled-back epochs under its local policy.

A holder without prior state retrieves the complete summary. A synchronized holder authenticates and applies $\Delta_e$. For $b$ changes and fixed~$k$, this authenticated delta transfers $\mathcal O(b)$ bytes and requires $\mathcal O(bd)$ local operations. A complete summary stores materialized nodes and revoked leaves, requiring $\mathcal O(N_{\mathrm{rev}}d)$ space, or $\mathcal O(N_{\mathrm{rev}})$ for fixed~$d$ (see Appendix Figure~\ref{fig:sp_verkle_summary}). \autoref{sec:discussion} covers trust on first use and offline root pinning. Synchronization remains separate from presentation as its metadata lies outside the unlinkability guarantee. 
\section{Evaluation Methodology}\label{sec:methodology}

RQ1 asks how holders can prove non-revocation without exposing credential-specific lookup information that could serve as stable handles. \autoref{sec:threat-model} defines the adversaries, assumptions, and security objectives, while \autoref{sec:implementation} presents the construction and security arguments. Functional validation complements these arguments with checks of forged paths, state reconstruction, and credential-status binding. Network synchronization metadata remains outside the cryptographic privacy guarantee. \autoref{sec:results-validation} reports the corresponding functional validation results.

RQ2 asks which proving, verification, and state-distribution costs arise from hiding credential-status lookups. The evaluation first compares SMT and Verkle under matched Groth16 conditions, then compares Groth16 and PLONK on the same Verkle construction. Additional experiments measure state acquisition and updates, mobile proving and verification, and cross-system proof costs. \autoref{sec:results-desktop}--\autoref{sec:results-context} report these quantitative results. 

\subsection{Functional Validation}
Functional validation addresses RQ1 at the implementation level and follows the malicious-holder cases from \autoref{sec:threat-model}. Five valid cases exercise non-revocation across all Verkle levels, including early termination at an authenticated empty child. Thirteen negative cases cover revocation, forged commitments and paths, invalid selectors, inconsistent secrets, mismatched public statements, and invalid KZG points. State tests require bootstrap reconstruction and incremental updates to reproduce each epoch root. Composition tests enforce a shared $VC_{\mathrm{id}}$ and reject mismatched identifiers, issuer keys, and signatures. The deployed Verkle service processes ten valid requests per proof backend and rejects malformed, unsupported, corrupted, and revoked requests. The tests provide implementation-level evidence for status correctness and credential-status binding. 

\subsection{Desktop Proof and Verification Costs}
RQ2 first measures proof and verification costs under controlled desktop conditions. A matched SMT--Verkle experiment compares registry backends over the same $2^{50}$-position address space. Both backends use the same circuit field, Groth16 proof system, hardware, and trial protocol. \textsc{ShadowPath}-SMT authenticates a depth-50 Poseidon path, while \textsc{ShadowPath}-Verkle authenticates five KZG levels with $k=1\,024$.

A second experiment compares Groth16 and PLONK on the same Verkle construction with $k=1024$ and $d=5$. A separate PLONK sweep varies $k\in\{256,512,1\,024\}$ and $d\in\{3,4,5\}$. Each configuration contains 1\,000 credentials and covers an address space between $2^{24}$ and $2^{50}$ positions. 

\noindent\textbf{Proof measurements.}
Measured quantities include witness construction, proof generation, verification, and proof size. Proof-generation timing starts when the selected proof backend is invoked and excludes compilation, setup, key loading, and witness construction. Witness construction is measured separately. Verification timing covers the backend proof check, and reported proof sizes exclude public inputs. Composition overhead follows the same Groth16 desktop protocol. 

The evaluation generates the Verkle KZG SRS and the Groth16/PLONK parameters locally using operating-system randomness. Setup and key derivation are excluded from all reported timings. Local parameter generation leaves the corresponding setup secrets known to the evaluator; these parameters are therefore restricted to evaluation. Deployment requires parameters generated through a sound ceremony. 

\noindent\textbf{Desktop protocol.}
Desktop experiments were run on an Apple M2 Pro with \qty{16}{\giga\byte} of RAM, with one configuration active at a time. Each configuration was preceded by two CPU-idle measurements above \qty{75}{\percent}, sampled \qty{30}{\second} apart. Each configuration comprised 30 sequential retained trials; no separate untimed warm-up was discarded. Medians and interquartile ranges summarize the repeated measurements. Single-proof verification latency does not capture verifier behavior when multiple presentations arrive concurrently. A separate microbenchmark evaluates the verification endpoint under parallel request load.

\noindent\textbf{Verifier load.}
Each proof backend processed 300 requests at concurrency 32, where concurrency denotes the number of verification requests processed in parallel. One run was executed per backend. The benchmark reused a single valid proof across requests, so a benchmark flag disabled normal challenge consumption. The measurements characterize verification-endpoint processing under concurrent load rather than end-to-end protocol throughput. Because each configuration was executed once, the experiment does not support estimates of between-run variance. Holder-local lookup requires authenticated registry state before proof generation. RQ2 measures the cost of acquiring and maintaining that state. 

\subsection{State-Distribution Costs}
State experiments distinguish bootstrap from synchronized operation. A bootstrapping holder reconstructs the registry from the complete published summary, whereas a synchronized holder retains prior state and applies authenticated deltas.

\noindent\textbf{Registry reconstruction and updates.}
The deployed Verkle experiment started with 50 revoked credentials and applied two five-entry updates, producing populations of 50, 55, and 60. A separate network-free experiment compared SMT and Verkle at $10^2$, $10^3$, and $10^4$ revocations with update batches of 1, 5, 20, and 100 entries. Network transfer and IPFS/IPNS resolution were excluded. Each configuration was repeated five times and summarized by the median. Every reconstructed or updated root was checked for correctness.

The SMT and Verkle registry backends use different native summary serializers. Their summary sizes serve as descriptive measurements. Separate Verkle growth measurements cover $10^2$--$10^5$ revocations. Larger values are linearly extrapolated from the measured uncompressed bytes per revoked entry at $10^5$. The largest measured registry population is additionally used to evaluate complete-summary transfer under controlled network conditions.

\noindent\textbf{Complete-summary transfer.}
Transfer was measured at \num{100000} revoked credentials using raw and gzip-compressed complete summaries. Thirty HTTP transfers were executed over a controlled link with a \qty{300}{\mega\bit\per\second} downlink, \qty{500}{\milli\second} round-trip time, \qty{40}{\milli\second} jitter, and \qty{0.001}{\percent} configured packet loss. Timing includes network transfer but excludes IPNS resolution, integrity checking, local reconstruction, and proof generation. The network profile represents a controlled high-latency stress condition rather than a measured cellular trace. 

\subsection{Mobile Proving Costs}
Desktop and state experiments characterize the costs of proof generation, verification, and state acquisition. This experiment evaluates proof generation on mobile devices that can run wallets. Primary measurements were taken using an iPhone 17 Pro and a Galaxy S25. Each device executed both Groth16--PLONK and PLONK--Groth16 orders with five trials per backend, followed by a \qty{60}{\second} cooldown after each proof. Artifacts and proving keys were cached before measurement. 

Eight additional devices followed the same protocol. Proof generation and verification were recorded together with memory and thermal telemetry. Memory values are not compared across operating systems because iOS and Android expose different metrics. The \qty{1}{\second} and \qty{10}{\second} levels are illustrative proof-generation thresholds, not measured usability limits. Proof timing excludes state retrieval, registry reconstruction, and the surrounding credential protocol. Geekbench~6 scores provide descriptive hardware context~\cite{geekbench6internals}. Mobile SMT proving and registry reconstruction are excluded from the evaluation. A final experiment compares the measured proof costs with those of related systems, without treating their protocols as equivalent.

\subsection{Prior Work Comparison}
Pinned revisions of zk-creds~\cite{zk-creds}, UPPR~\cite{11264637}, and zkRevoke~\cite{manimaran2026zkRevoke} were measured independently on the same desktop host. The same CPU-idle criterion and 30-trial protocol were applied. Compilation, setup, key loading, and witness construction preceded timing, while fresh clones and clean builds checked build-level reproducibility. ZEBRA~\cite{rathee2022zebra} contributes source-reported measurements and was not reproduced on the evaluation host. Because the compared systems prove different statements and use different constructions, the measurements provide proof-cost context rather than a normalized protocol ranking. \autoref{sec:results-context} reports the comparison.

\section{Experimental Results and Analysis}
\label{sec:experiments}

Section~\ref{sec:methodology} defines the evaluation protocol. Functional validation addresses RQ1. Desktop, state-distribution, mobile, and prior-work results address RQ2.

\subsection{Functional Validation}
\label{sec:results-validation}

The Verkle implementation accepted all five valid witnesses, one at each termination level, and rejected all 13 negative cases. Rejections covered revocation, forged commitments and path mutations, invalid selectors, inconsistent secrets, mismatched public statements, and invalid KZG points. The deployed service additionally completed ten valid requests per proof backend and rejected malformed, unsupported, corrupted, or revoked requests. Bootstrap reconstruction and incremental updates reproduced the expected authenticated roots; Section~\ref{sec:results-state} reports the corresponding state costs.

The composition proof of concept accepted the shared-$VC_{\mathrm{id}}$ case and rejected a different credential identifier, a mismatched issuer key, and a tampered signature. These outcomes provide implementation-level evidence for status correctness and credential-status binding. Cryptographic soundness and privacy are addressed by Section~\ref{sec:implementation}. The remaining results address RQ2, beginning with controlled desktop proof and verification costs.

\begin{tcolorbox}[
colback=gray!7,
coltitle=black,
frame hidden,
sharp corners,
enhanced,
left=2mm,
boxrule=0pt,
borderline west={0.75pt}{0pt}{snet-red}
]
\small
\textbf{\textsc{Takeaway 1:}} Functional validation accepted all five valid cases, rejected all 13 negative cases, and reproduced the expected epoch roots. The results provide implementation-level evidence for status correctness and credential-status binding.
\end{tcolorbox}

\subsection{Desktop Proof and Verification Costs}
\label{sec:results-desktop}

\begin{table}[!t]
\centering
\caption{Measured proof and verification costs (30-trial medians).}
\label{tab:proof-side-costs}
\footnotesize
\setlength{\tabcolsep}{2pt}
\renewcommand{\arraystretch}{1.04}
\begin{tabular*}{\columnwidth}{@{\extracolsep{\fill}}l
    S[table-format=5.1,group-digits=false]
    S[table-format=2.2]
    S[table-format=4.0,group-digits=false]@{}}
\toprule
\textbf{Relation} &
\multicolumn{1}{c}{\textbf{Proving (\unit{\milli\second})}} &
\multicolumn{1}{c}{\textbf{Verification (\unit{\milli\second})}} &
\multicolumn{1}{c}{\textbf{Proof (\unit{\byte})}} \\
\midrule

\multicolumn{4}{@{}l}{Groth16}\\[-0.3ex]
\quad\textsc{ShadowPath}-SMT
    & 371.6 & 3.70 & 388 \\
\quad\textsc{ShadowPath}-Verkle
    & 2109.5 & 7.55 & 484 \\

\addlinespace[0.3ex]
\multicolumn{4}{@{}l}{PLONK}\\[-0.3ex]
\quad\textsc{ShadowPath}-Verkle
    & 25619.8 & 9.60 & 1352 \\

\bottomrule
\end{tabular*}

\vspace{0.8mm}
\parbox{\columnwidth}{\footnotesize\raggedright
The root-linked \textsc{ShadowPath} relations share a $2^{50}$-position address space: SMT uses $k=2,d=50$, whereas Verkle uses $k=1024,d=5$. The matched SMT--Verkle comparison fixes Groth16; PLONK is evaluated for Verkle. Proof bytes exclude public inputs.}
\end{table}

Table~\ref{tab:proof-side-costs} reports the matched SMT--Verkle comparison. Under Groth16, median proving times were \qty{371.6}{\milli\second} for \textsc{ShadowPath}-SMT and \qty{2109.5}{\milli\second} for \textsc{ShadowPath}-Verkle, a $5.68\times$ difference. Median verification times were \qty{3.70}{\milli\second} and \qty{7.55}{\milli\second}, respectively, a $2.04\times$ difference. Verification-time IQRs were \qty{0.09}{\milli\second} for SMT and \qty{0.17}{\milli\second} for Verkle.

The Groth16 proof measured \qty{388}{\byte} for \textsc{ShadowPath}-SMT and \qty{484}{\byte} for \textsc{ShadowPath}-Verkle. The additional \qty{96}{\byte} reflects one extra gnark prover commitment introduced by the deferred emulated-field checks. The corresponding BW6-761 $\mathbb{G}_1$ element affects both proof serialization and verifier work. The verification difference therefore cannot be attributed solely to circuit size.

For \textsc{ShadowPath}-Verkle, median proving increased from \qty{2109.5}{\milli\second} with Groth16 to \qty{25619.8}{\milli\second} with PLONK, a $12.14\times$ difference. Median verification increased from \qty{7.55}{\milli\second} to \qty{9.60}{\milli\second}, while proof size increased from \qty{484}{\byte} to \qty{1352}{\byte} ($2.79\times$). Proving-time IQRs were \qty{74.5}{\milli\second} and \qty{1202.9}{\milli\second}, respectively.
Adding credential authentication to the Verkle status relation increased the R1CS from \num{74649} to \num{86034} constraints (\qty{15.3}{\percent}). Across 30 Groth16 trials, median proving and verification times were \qty{2.207}{\second} (IQR \qty{77.6}{\milli\second}) and \qty{7.645}{\milli\second} (IQR \qty{0.210}{\milli\second}), respectively. Proving latency was \qty{4.6}{\percent} above the standalone Verkle relation.

\begin{figure}[!t]
\centering
\includegraphics[width=\columnwidth]{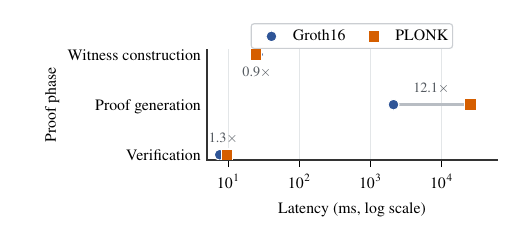}
\caption{\textsc{ShadowPath}-Verkle latency by proof phase at $k=1\,024$ and $d=5$. Labels show the PLONK-to-Groth16 ratio for each phase; the latency axis is logarithmic.}
\label{fig:component_profiling}
\end{figure}

Figure~\ref{fig:component_profiling} shows that proof generation dominates the holder-side computation. Median witness construction required \qty{26.1}{\milli\second} with Groth16 and \qty{24.5}{\milli\second} with PLONK, whereas proof generation required \qty{2109.5}{\milli\second} and \qty{25619.8}{\milli\second}, respectively. Proof generation accounted for more than \qty{98}{\percent} of the combined witness-construction and proof-generation latency for both proof backends.

\begin{figure}[t]
\centering
\includegraphics[width=0.82\columnwidth]{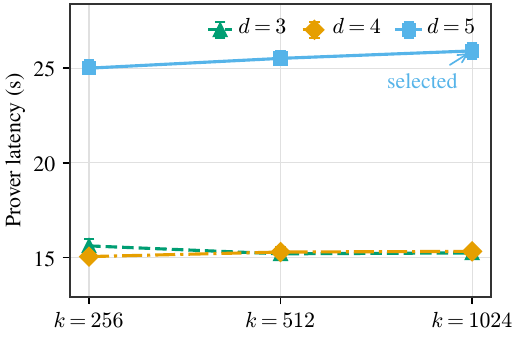}
\caption{PLONK proving latency across Verkle geometries. The annotation marks the selected $2^{50}$-position profile at $k=1\,024$ and $d=5$.}
\label{fig:param_sweep}
\end{figure}

Figure~\ref{fig:param_sweep} isolates the effect of Verkle geometry on PLONK proving. Across nine configurations, median proving times ranged from \qty{15.05}{\second} to \qty{25.92}{\second}. Depths three and four remained within \qtyrange{15.05}{15.62}{\second}, whereas every depth-five configuration required at least \qty{25.01}{\second}. At depth five, the circuit crosses the next PLONK evaluation domain and the proving key grows from \qty{50.5}{\mega\byte} to \qty{100.8}{\mega\byte}. Branching factor had a smaller, non-monotonic effect at fixed depth.

The selected $k=1\,024,d=5$ profile required \qty{25.92}{\second} in the geometry sweep and \qty{25.62}{\second} in the independent measurement. The profile was selected because five levels cover the matched $2^{50}$-position address space, not because it minimizes proving latency. At that address space, the Verkle relation contains \num{74649} constraints, compared with \num{15303} for SMT. Five independently verified KZG openings therefore did not reduce proof-generation cost relative to 50 Poseidon levels in the evaluated construction. Different geometry or aggregated openings could change this ordering.

Verifier-side processing remained substantially faster than proof generation. At concurrency 32, the 300-request runs sustained \num{346.8} accepted requests per second with Groth16 and \num{284.1} with PLONK. Median and 95th-percentile request latencies were \qty{73.2}{\milli\second} and \qty{214.07}{\milli\second} for Groth16, and \qty{99.54}{\milli\second} and \qty{225.32}{\milli\second} for PLONK. Because each backend was measured once at this concurrency level, the result characterizes endpoint behavior but does not support between-run variance estimates.

\begin{tcolorbox}[
colback=gray!7,
coltitle=black,
frame hidden,
sharp corners,
enhanced,
left=2mm,
boxrule=0pt,
borderline west={0.75pt}{0pt}{snet-red}
]
\small
\textbf{\textsc{Takeaway 2:}} Under matched Groth16 conditions, \textsc{ShadowPath}-Verkle required $5.68\times$ the proving latency of \textsc{ShadowPath}-SMT despite reducing path depth from 50 hash levels to five KZG levels. Within Verkle, PLONK required $12.14\times$ the proving latency of Groth16.
\end{tcolorbox}

\subsection{State-Distribution Costs}
\label{sec:results-state}

\begin{figure}[t]
\centering
\includegraphics[width=0.82\columnwidth]{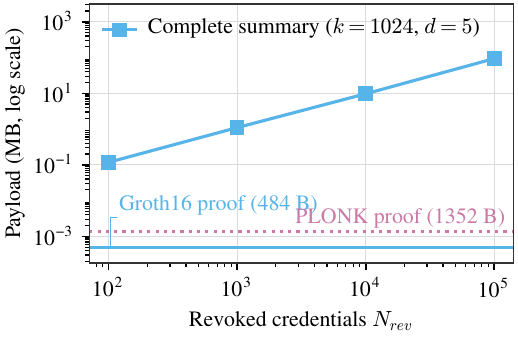}
\caption{Complete-summary growth for \textsc{ShadowPath}-Verkle at $k=1\,024$ and $d=5$. Summary size grows with the number of revoked credentials, whereas the Groth16 and PLONK proof sizes remain fixed at these circuit parameters.}
\label{fig:registry_growth_comm}
\end{figure}

State-acquisition cost separates bootstrap or recovery from synchronized operation. Figure~\ref{fig:registry_growth_comm} shows that the Verkle complete summary grows from \qty{0.118}{\mega\byte} at $10^2$ revocations to \qty{93.95}{\mega\byte} at $10^5$. Under the controlled high-latency profile, 30 raw-summary transfers yielded a median of \qty{24.86}{\second} (IQR \qty{10.08}{\second}). Gzip reduced the payload to \qty{38.09}{\mega\byte} and the median transfer time to \qty{13.76}{\second} (IQR \qty{8.22}{\second}).

Synchronized operation required less transfer. At 50 revocations, the complete summary occupied \qty{48459}{\byte}, whereas each subsequent five-update authenticated delta occupied \qty{1385}{\byte}. Both updates therefore transferred \qty{2770}{\byte} across two epochs and reproduced the corresponding authenticated roots. Linear extrapolation from the measured uncompressed bytes per revoked entry at $10^5$ gives approximately \qty{0.94}{\giga\byte} at $10^6$ revocations and \qty{9.4}{\giga\byte} at $10^7$. These values are projections for bootstrap or recovery, not measured per-epoch transfer.

The network-free comparison isolates local state processing. All 24 combinations of registry backend, population, and update batch reproduced their expected roots. At $10^4$ revocations, the native complete-summary encodings occupied \qty{7.82}{\mega\byte} for Verkle and \qty{43.75}{\mega\byte} for SMT. Different serializers preclude a comparison of normalized storage. Reconstruction required \qty{427.18}{\second} for Verkle and \qty{23.44}{\second} for SMT. Their common authenticated-delta format produced payloads of \qtyrange{400}{7033}{\byte} for batches of 1--100 updates; applying and verifying 100 updates required \qty{6.77}{\second} for Verkle and \qty{240.53}{\milli\second} for SMT.

The measurements therefore separate registry-sized bootstrap or recovery costs from update-batch-sized costs during synchronized operation. Proof size remains independent of registry population at fixed circuit parameters.

\begin{tcolorbox}[
colback=gray!7,
coltitle=black,
frame hidden,
sharp corners,
enhanced,
left=2mm,
boxrule=0pt,
borderline west={0.75pt}{0pt}{snet-red}
]
\small
\textbf{\textsc{Takeaway 3:}} Synchronized holders receive update-batch-sized authenticated deltas, whereas bootstrap and recovery require the complete summary. At $10^4$ revocations, Verkle reconstruction required \qty{427.18}{\second} versus \qty{23.44}{\second} for SMT.
\end{tcolorbox}

\subsection{Mobile Proving Costs}
\label{sec:results-mobile}

\begin{table}[t]
\centering
\caption{\textsc{ShadowPath}-Verkle latency on the two primary mobile devices.}
\label{tab:physical-device-results}
\footnotesize
\setlength{\tabcolsep}{3.2pt}
\renewcommand{\arraystretch}{1.10}
\begin{tabular*}{\columnwidth}{@{\extracolsep{\fill}}lcc@{}}
\toprule
\textbf{Backend}
&
\textbf{Proving $\tilde{\mu}$ [range] (\unit{\second})}
&
\textbf{Verification $\tilde{\mu}$ (\unit{\milli\second})}
\\
\midrule
\multicolumn{3}{@{}l}{\textit{iPhone 17 Pro (iOS 26.6)}} \\
Groth16 & \num{3.11}\,[\num{3.05}--\num{3.23}] & \num{5.31} \\
PLONK   & \num{39.12}\,[\num{38.43}--\num{42.19}] & \num{8.08} \\
\cmidrule{1-3}
\multicolumn{3}{@{}l}{\textit{Galaxy S25 (Android 16)}} \\
Groth16 & \num{2.93}\,[\num{2.88}--\num{3.36}] & \num{19.08} \\
PLONK   & \num{55.29}\,[\num{43.73}--\num{63.98}] & \num{17.05} \\
\bottomrule
\end{tabular*}
\parbox{\columnwidth}{\footnotesize\raggedright
\vspace{1.5mm}
Brackets denote observed proving ranges. Each row pools ten
trials across both counterbalanced backend orders. Appendix Table~\ref{tab:mobile_device_matrix} reports the results for all tested devices.}
\end{table}

On the two primary devices, median Groth16 proving required \qty{3.11}{\second} on the iPhone 17 Pro and \qty{2.93}{\second} on the Galaxy S25. Both exceeded the \qty{1}{\second} threshold but remained below \qty{10}{\second}. PLONK required \qty{39.12}{\second} and \qty{55.29}{\second}, respectively, exceeding \qty{10}{\second} on both devices. Table~\ref{tab:physical-device-results} reports the underlying 40 trials.

\begin{figure}[!t]
\centering
\begingroup
\begin{tikzpicture}[
  x=0.58cm,
  y=0.60cm,
  font=\small,
  ios/.style={blue!70!black},
  android/.style={gray},
  annotation/.style={
    font=\footnotesize,
    align=left,
    rounded corners=1.5pt,
    draw=black!20,
    fill=white,
    inner xsep=2.5pt,
    inner ysep=1.5pt
  },
  leader/.style={draw=black!28,thin}
]


\draw[black!14,thin] (1.081,0) -- (1.081,7);
\draw[black!14,thin] (2.883,0) -- (2.883,7);
\draw[black!14,thin] (4.685,0) -- (4.685,7);
\draw[black!14,thin] (6.486,0) -- (6.486,7);
\draw[black!14,thin] (8.288,0) -- (8.288,7);
\draw[black!14,thin] (10.090,0) -- (10.090,7);
\draw[black!14,thin] (11.892,0) -- (11.892,7);

\node[anchor=north] at (1.081,-0.12) {1\,000};
\node[anchor=north] at (2.883,-0.12) {1\,500};
\node[anchor=north] at (4.685,-0.12) {2\,000};
\node[anchor=north] at (6.486,-0.12) {2\,500};
\node[anchor=north] at (8.288,-0.12) {3\,000};
\node[anchor=north] at (10.090,-0.12) {3\,500};
\node[anchor=north] at (11.892,-0.12) {4\,000};

\draw[black!14,thin] (0,0.000) -- (12,0.000);
\draw[black!8,thin] (0,0.916) -- (12,0.916);
\draw[black!8,thin] (0,1.44) -- (12,1.44);
\draw[black!8,thin] (0,1.832) -- (12,1.832);
\draw[black!8,thin] (0,2.126) -- (12,2.126);
\draw[black!8,thin] (0,2.367) -- (12,2.367);
\draw[black!8,thin] (0,2.571) -- (12,2.571);
\draw[black!8,thin] (0,2.75) -- (12,2.75);
\draw[black!8,thin] (0,2.91) -- (12,2.91);
\draw[black!14,thin] (0,3.05) -- (12,3.05);
\draw[black!8,thin] (0,3.958) -- (12,3.958);
\draw[black!8,thin] (0,4.5) -- (12,4.5);
\draw[black!8,thin] (0,4.88) -- (12,4.88);
\draw[black!8,thin] (0,5.17) -- (12,5.17);
\draw[black!8,thin] (0,5.409) -- (12,5.409);
\draw[black!8,thin] (0,5.613) -- (12,5.613);
\draw[black!8,thin] (0,5.789) -- (12,5.789);
\draw[black!8,thin] (0,5.945) -- (12,5.945);
\draw[black!14,thin] (0,6.084) -- (12,6.084);

\node[anchor=east] at (-0.15,0.000) {$10^3$};
\node[anchor=east] at (-0.15,3.042) {$10^4$};
\node[anchor=east] at (-0.15,6.084) {$10^5$};

\draw[black,thin] (0,0) -- (12,0);
\draw[black,thin] (0,0) -- (0,7);

\node[anchor=north] at (6.01,-0.9)
  {representative Geekbench 6 single-core scores};

\node[rotate=90,anchor=south,xscale=1.1,yscale=1.1]
  at (-1.19,3.11)
  {Proof-generation latency (ms, log scale)};

\path[fill=blue!12,draw=none]
  (5.121,2.193) -- (6.317,2.006) -- (7.200,1.864) --
  (7.312,1.839) -- (8.058,1.838) -- (10.645,1.497) --
  (11.258,1.472) -- (11.258,1.547) -- (10.645,1.616) --
  (8.058,2.047) -- (7.312,1.872) --
  (7.200,2.153) -- (6.317,2.356) -- (5.121,2.615) -- cycle;

\path[fill=blue!12,draw=none]
  (5.121,5.658) -- (6.317,5.296) -- (7.200,5.223) --
  (7.312,5.142) -- (8.058,5.138) -- (10.645,4.883) --
  (11.258,4.821) -- (11.258,4.944) -- (10.645,5.080) --
  (8.058,5.578) -- (7.312,5.279) --
  (7.200,5.341) -- (6.317,5.679) -- (5.121,6.024) -- cycle;

\path[fill=gray!15,draw=none]
  (1.128,3.026) -- (8.328,1.459) -- (8.670,1.396) --
  (8.670,1.601) -- (8.328,1.522) -- (1.128,3.115) -- cycle;

\path[fill=gray!15,draw=none]
  (1.128,6.300) -- (8.328,4.814) -- (8.670,4.992) --
  (8.670,5.494) -- (8.328,5.153) -- (1.128,6.822) -- cycle;

\draw[ios,thick]
  (5.121,2.214) -- (6.317,2.017) -- (7.200,1.879) --
  (7.312,1.853) -- (8.058,1.905) -- (10.645,1.521) --
  (11.258,1.497);

\fill[ios] (5.121,2.214) circle (2.0pt);
\fill[ios] (6.317,2.017) circle (2.0pt);
\fill[ios] (7.200,1.879) circle (2.0pt);
\fill[ios] (7.312,1.853) circle (2.0pt);
\fill[ios] (8.058,1.905) circle (2.0pt);
\fill[ios] (10.645,1.521) circle (2.0pt);
\fill[ios] (11.258,1.497) circle (2.0pt);

\draw[ios,thick]
  (5.121,5.817) -- (6.317,5.492) -- (7.200,5.245) --
  (7.312,5.202) -- (8.058,5.183) -- (10.645,4.927) --
  (11.258,4.844);

\path[fill=blue!70!black,draw=blue!70!black]
  (5.121,5.907) -- (5.041,5.747) -- (5.201,5.747) -- cycle;
\path[fill=blue!70!black,draw=blue!70!black]
  (6.317,5.582) -- (6.237,5.422) -- (6.397,5.422) -- cycle;
\path[fill=blue!70!black,draw=blue!70!black]
  (7.200,5.335) -- (7.120,5.175) -- (7.280,5.175) -- cycle;
\path[fill=blue!70!black,draw=blue!70!black]
  (7.312,5.292) -- (7.232,5.132) -- (7.392,5.132) -- cycle;
\path[fill=blue!70!black,draw=blue!70!black]
  (8.058,5.273) -- (7.978,5.113) -- (8.138,5.113) -- cycle;
\path[fill=blue!70!black,draw=blue!70!black]
  (10.645,5.017) -- (10.565,4.857) -- (10.725,4.857) -- cycle;
\path[fill=blue!70!black,draw=blue!70!black]
  (11.258,4.934) -- (11.178,4.774) -- (11.338,4.774) -- cycle;

\draw[android,dashed,thick]
  (1.128,3.098) -- (8.328,1.494) -- (8.670,1.420);

\fill[android] (1.128,3.098) circle (2.0pt);
\fill[android] (8.328,1.494) circle (2.0pt);
\fill[android] (8.670,1.420) circle (2.0pt);

\draw[android,dashed,thick]
  (1.128,6.378) -- (8.328,5.005) -- (8.670,5.301);

\path[fill=gray,draw=gray]
  (1.128,6.468) -- (1.048,6.308) -- (1.208,6.308) -- cycle;
\path[fill=gray,draw=gray]
  (8.328,5.095) -- (8.248,4.935) -- (8.408,4.935) -- cycle;
\path[fill=gray,draw=gray]
  (8.670,5.391) -- (8.590,5.231) -- (8.750,5.231) -- cycle;

\node[annotation,text=gray,anchor=west,xscale=1.1,yscale=1.1]
  (SMA546B) at (0.14,5.06) {Galaxy A54};
\draw[leader,thick] (SMA546B.south) -- (1.128,3.098);

\node[annotation,text=gray,anchor=east,xscale=1.1,yscale=1.1]
  (SMS938B) at (5.32,0.77) {Galaxy S25 Ultra};
\draw[leader,thick] (SMS938B.north) -- (8.328,1.494);

\node[annotation,text=gray,anchor=west,xscale=1.1,yscale=1.1]
  (SMS931B) at (5.89,0.75) {Galaxy S25};
\draw[leader,thick] (SMS931B.north) -- (8.670,1.420);

\node[annotation,text=blue!70!black,anchor=west,xscale=1.1,yscale=1.1]
  (iPhone132) at (2.03,3.61) {iPhone 12};
\draw[leader,thick] (iPhone132.south) -- (5.121,2.214);

\node[annotation,text=blue!70!black,anchor=west,xscale=1.1,yscale=1.1]
  (iPhone142) at (3.52,4.58) {iPhone 13 Pro};
\draw[leader,thick] (iPhone142.south) -- (6.317,2.017);

\node[annotation,text=blue!70!black,anchor=west,xscale=1.1,yscale=1.1]
  (iPhone155) at (5.04,3.65) {iPhone 15 Plus};
\draw[leader,thick] (iPhone155.south) -- (7.200,1.879);

\node[annotation,text=blue!70!black,anchor=west,xscale=1.1,yscale=1.1]
  (iPhone153) at (7.43,4.41) {iPhone 14 Pro Max};
\draw[leader,thick] (iPhone153.south) -- (7.312,1.853);

\node[annotation,text=blue!70!black,anchor=west,xscale=1.1,yscale=1.1]
  (iPhone162) at (7.64,3.00) {iPhone 15 Pro Max};
\draw[leader,thick] (iPhone162.south) -- (8.058,1.905);

\node[annotation,text=blue!70!black,anchor=east,xscale=1.1,yscale=1.1]
  (iPhone183) at (11.88,0.75) {iPhone 17};
\draw[leader,thick] (iPhone183.north) -- (10.645,1.521);

\node[annotation,text=blue!70!black,anchor=east,xscale=1.1,yscale=1.1]
  (iPhone181) at (12.68,2.25) {iPhone 17 Pro};
\draw[leader,thick] (iPhone181.south) -- (11.258,1.497);


\node[anchor=east,font=\footnotesize] at (6.92,8.2) {OS:};

\draw[ios,thick] (7.09,8.2) -- (7.69,8.2);
\node[anchor=west] at (7.79,8.2) {iOS};

\draw[android,dashed,thick] (9.15,8.2) -- (9.75,8.2);
\node[anchor=west] at (9.85,8.2) {Android};

\node[anchor=east,font=\footnotesize] at (6.81,7.56) {Backend:};

\fill[black] (7.02,7.54) circle (2.0pt);
\node[anchor=west] at (7.22,7.54) {Groth16};

\draw[fill=black, black]
  (9.7,7.6) -- (9.62,7.44) -- (9.78,7.44) -- cycle;
\node[anchor=west] at (9.9,7.52) {PLONK};

\end{tikzpicture}
\endgroup
\caption{Proof-generation latency for \textsc{ShadowPath}-Verkle across ten tested mobile devices. Points show pooled ten-trial medians per device and proof backend. Shaded bands show observed min--max ranges. Solid and dashed lines distinguish iOS and Android, while circles and triangles denote Groth16 and PLONK. The \qty{1}{\second} and \qty{10}{\second} levels mark illustrative proof-generation thresholds. Appendix Table~\ref{tab:mobile_device_matrix} reports the complete measurements.}
\label{fig:mobile_soc_pareto}
\end{figure}
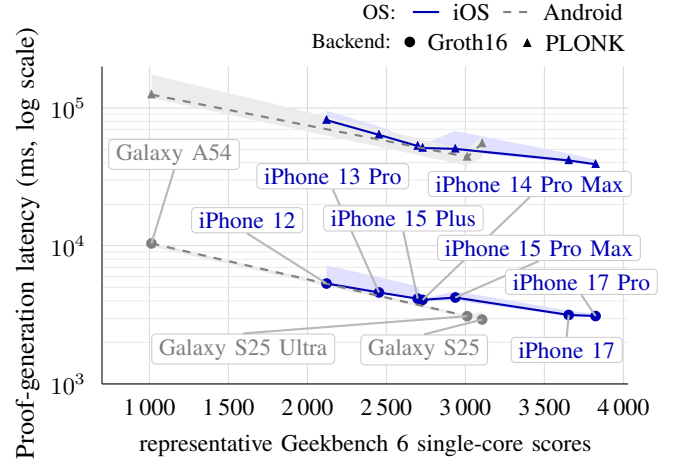

Across ten physical devices, 200 \textsc{ShadowPath}-Verkle trials were completed. Groth16 remained below \qty{10}{\second} on nine devices. The Galaxy A54 reached \qty{10.46}{\second}. No device remained below \qty{1}{\second}, and PLONK exceeded \qty{10}{\second} on every device. Figure~\ref{fig:mobile_soc_pareto} summarizes the results.
PLONK runs also showed higher resource-use indicators on several devices. Peak OS-reported PLONK memory use reached approximately \qty{5.0}{\giga\byte}, and several runs reached serious or severe thermal states. iOS application footprint and Android resident-set size use different metrics, so memory values are not compared across operating systems. Mobile SMT proving and holder-side state reconstruction were not evaluated.

\begin{tcolorbox}[
colback=gray!7,
coltitle=black,
frame hidden,
sharp corners,
enhanced,
left=2mm,
boxrule=0pt,
borderline west={0.75pt}{0pt}{snet-red}
]
\small
\textbf{\textsc{Takeaway 4:}} Mobile proving is viable for attended credential presentations, but current proving costs remain incompatible with latency-sensitive interactions. Proof-system choice therefore directly constrains the interaction model.
\end{tcolorbox}

\subsection{Prior Work Comparison}
\label{sec:results-context}

\begin{figure*}[!t]
\centering
\includegraphics[width=0.85\textwidth]{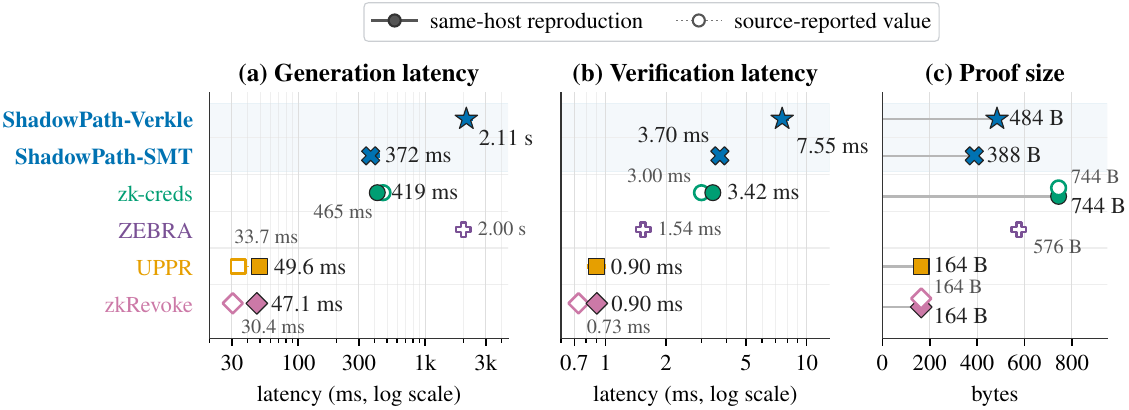}
\caption{Proof-cost context for \textsc{ShadowPath} and related artifacts. Filled markers denote same-host measurements; open markers denote source-reported values. ZEBRA was not reproduced on the evaluation host, so all ZEBRA values shown are source-reported: \qty{2}{\second} proof generation, a \qty{576}{\byte} proof, and an estimated \qty{1.54}{\milli\second} verification time~\cite{rathee2022zebra}. UPPR provides no comparable source-reported verification latency or serialized proof size. This figure only provides cost context as the systems prove different statements.} 
\label{fig:comparison}
\end{figure*}

Figure~\ref{fig:comparison} compares proof costs with related artifacts. 
Different proof statements preclude a normalized ranking. Among the same-host measurements, \textsc{ShadowPath}-Verkle had the highest proving and verification latencies. \textsc{ShadowPath}-SMT proved in \qty{371.6}{\milli\second} versus \qty{419}{\milli\second} for zk-creds and produced a \qty{388}{\byte} versus \qty{744}{\byte} proof. UPPR and zkRevoke were faster and smaller, but prove different token-based statements.

\begin{tcolorbox}[
colback=gray!7,
coltitle=black,
frame hidden,
sharp corners,
enhanced,
left=2mm,
boxrule=0pt,
borderline west={0.75pt}{0pt}{snet-red}
]
\small
\textbf{\textsc{Takeaway 5:}} \textsc{ShadowPath}-SMT achieved proof-generation cost close to zk-creds while producing a smaller proof; \textsc{ShadowPath}-Verkle was the most expensive same-host design. UPPR and zkRevoke were faster and smaller but prove different token-based statements, so the comparison is contextual rather than a normalized ranking.
\end{tcolorbox}
\section{Discussion}
\label{sec:discussion}

\subsection{Deployment and Security Boundaries}
The EUDI Wallet Architecture and Reference Framework carries a status-list URL and credential-specific identifier or index~\cite[Sec.~6.6.3.7]{eudi_arf}. \textsc{ShadowPath} targets this lookup-privacy boundary; it is neither a drop-in ARF profile nor evidence of regulatory compliance. Similar needs arise in national wallets~\cite{bhutanNDI,utahVDC}.

The Verkle proof of concept composes status with an EdDSA credential relation. Production composition must preserve the same hidden identifier and exact session binding and derive the session value with a formally analyzed PRF keyed by a credential-bound holder secret; its circuit cost remains unmeasured. Fresh, consumed challenges prevent replay, and issuer--verifier collusion remains excluded.

\textsc{ShadowPath} does not guarantee complete, timely, or globally consistent publication. First-use and offline verifiers lack a trusted reference, and IPNS cannot exclude split views. Out-of-band pinning, transparency-published epoch heads, or anonymous retrieval could strengthen consistency or metadata privacy but remain unevaluated.

\subsection{Backend Trade-offs and Evaluation Scope}
At the matched $2^{50}$ point, SMT had lower proving, verification, local-processing, and proof-size costs than the evaluated Verkle construction. Five independently verified KZG openings did not offset the cost advantage of the 50-level Poseidon path. This artifact ordering may change with geometry, opening aggregation, or hashes such as Poseidon2~\cite{grassi2023poseidon2}. Both proofs are fixed-size at compiled parameters, while recovery requires reconstruction from a complete summary. Incompatible encodings preclude a normalized size comparison.

The primary mobile-device study measures Verkle proof-generation cost: Groth16 remained below the illustrative \qty{10}{\second} threshold but exceeded \qty{1}{\second}, while PLONK exceeded \qty{10}{\second} on both devices. Mobile SMT and state processing were not evaluated, so these results do not establish end-to-end mobile feasibility. Aggregated openings or circuit-friendly commitments could reduce Verkle costs; authenticated sharding or checkpoints could bound recovery time. These designs remain unimplemented and may add assumptions.

Wallets can apply authenticated deltas and cache witnesses before presentation, but fresh challenges prevent complete-proof precomputation. Delegated proving exposes private witness material unless the prover is trusted or isolated. 

Status mechanisms shift cost among downloads, witness maintenance, and holder proving~\cite{W3C_Bitstring2024,7958597,camenisch2002dynamic,10.1145/3634737.3637641,iden3,zk-creds,rathee2022zebra}. Comparisons must therefore state the proved relation and report lifecycle costs separately. We omit Verkle aggregation, mobile reconstruction, Internet-scale availability, wallet integration, and alternative synchronization.
\section{Conclusion}\label{sec:conclusion}
\textsc{ShadowPath} hides the credential registry lookup during presentation. It separates credential state distribution from credential-specific verification. The holder reconstructs the witness locally and proves non-revocation under a verifier-selected epoch root while hiding lookup metadata. This provides lookup/path privacy and issuer-query unobservability because no credential-specific status query reaches the issuer. Fresh session values provide presentation unlinkability by preventing verifier-visible status data from linking separate presentations. Together, these properties realize the three privacy objectives under the stated threat model.

Three lessons emerge from the evaluation. First, hiding the lookup inside a ZKP addresses only one part of the privacy problem. State acquisition and witness maintenance must also avoid credential-specific information. Second, shorter authentication paths do not necessarily reduce proof cost. The SMT backend was substantially more efficient than Verkle, despite using 10 times as many tree levels. Third, holder-side Groth16 proving is feasible on current mobile devices, although registry and proof backends strongly affect performance. Credential-status mechanisms should therefore be evaluated across the full lifecycle. \textsc{ShadowPath} characterizes these trade-offs and provides an empirical basis for future lookup-private credential-status systems.

Future research should extend the integration of private proving and distributed authenticated state. Aggregated openings, alternative bilinear-pairing constructions, multiparty computation, and circuit-friendly commitments may reduce proof costs while preserving distributed authenticated state. Post-quantum-secure commitments and proof systems provide a complementary direction beyond pairing-based constructions. Future work should also evaluate mobile device proving and integrate privacy-preserving synchronization mechanisms to reduce computational costs and leakage of network metadata.

\section*{Acknowledgment}
We thank from the Technische Universität Berlin, Thomas Cory for his valuable feedback and support for the Figures. This work was conducted at the SNET research group, which is part of T-Labs -- a public-private partnership between Technische Universität Berlin and Deutsche Telekom AG. 

\section*{Ethics Considerations}

\begingroup
\setlength{\parindent}{0pt}

Following the Menlo principles~\cite{bailey2012menlo}, we consider the ethical implications of conducting, publishing, and potentially deploying \textsc{ShadowPath}. The relevant stakeholders and their principal benefits, harms, and risks are: 

\begin{itemize}[leftmargin=*, itemsep=0.4mm, parsep=0pt]
    \item \textbf{Holders} may benefit from reduced status-layer linkability. Still, they bear the costs of computation, storage, and synchronization and may experience delays or exclusion when policy-fresh state or timely proof generation is unavailable.
    
    \item \textbf{Verifiers} gain locally verifiable status evidence without presentation-time issuer queries, but they bear the consequences of false acceptance, false rejection, and unavailable or stale state.
    
    \item \textbf{Issuers and registry operators} avoid credential-specific presentation queries, but they must publish authentic and available state; visible publication or synchronization interactions may still reveal metadata.
    
    \item The \textbf{general public} may benefit from less traceable credential verification and reliable revocation, but may bear harms from credential fraud, unjustified exclusion, service disruption, or weakened accountability.
    
    \item The \textbf{research community} gains auditable evidence, implementation artifacts, and reproducibility material, but premature or uncritical reuse of an experimental system could propagate design or implementation weaknesses.
\end{itemize}


\noindent\textbf{Benefits, Harms, and Respect for Persons.} People who present credentials may benefit when status checks reveal less information that issuers or verifiers could use to link their activities across contexts. This can support privacy and autonomy, particularly when using credentials reveals sensitive aspects of a person's life. However, people and organizations that rely on credential decisions also need timely, accurate revocation information. Stale or inconsistent state, implementation errors, compromised setup parameters, or unavailable updates could cause false acceptance or false rejection. False acceptance may expose others to fraud, safety risks, or financial harm, whereas false rejection may wrongfully deny an honest holder access to a service. Verifiers should apply risk-proportionate freshness policies and distinguish unavailable status from confirmed revocation.

Within its stated assumptions, \textsc{ShadowPath} authenticates non-revocation under a verifier-selected epoch root and binds the result to the same hidden credential and presentation session. It does not, however, ensure complete, timely, or globally consistent publication of registry state. Reduced status-layer observability may also limit legitimate investigations of misuse despite the intended privacy benefit.

Credential verification may affect access to employment, education, travel, financial services, or other important areas of everyday life. Deployments should explain what presentation and synchronization metadata remains observable, how transcripts are retained, and which parties may collude. They should also provide authenticated alternative procedures during a transition period, as well as bounded retry, recovery, and appeal mechanisms when policy-fresh state cannot be obtained or a device cannot complete a proof in time. A failed or delayed proof must not by itself be treated as holder misconduct.


\noindent\textbf{Justice.} Holder-side proving costs are not distributed equally. A device's computational resources, connectivity, battery capacity, accessibility features, and available technical support may determine whether a person can complete verification within the required time. Holders with constrained devices, unreliable or expensive connectivity, accessibility needs, or limited technical support may experience disproportionate delay or exclusion. Our evaluation identifies this risk, but neither estimates the affected population nor evaluates usability.

Future studies should include older and entry-level devices, varied network conditions, accessibility requirements, and application-specific latency budgets. Deployments should offer authenticated, privacy-preserving alternatives so that access does not depend on owning high-end hardware.


\noindent\textbf{Respect for Law and Public Interest.} Societal, ethical, and legal responsibility is shared among the identified stakeholders to ensure that privacy protection does not undermine accountability or applicable legal obligations. Credential-specific status queries can allow issuers, verifiers, or network observers to link a person's activities across contexts. \textsc{ShadowPath} reduces this risk by moving credential-specific lookup to the holder's device. Fresh challenges and challenge consumption prevent direct replay, while the presentation randomizer avoids a deterministic transcript handle under the stated session-value pseudorandomness assumption. This protection does not cover issuer--verifier collusion.

State synchronization remains outside the cryptographic privacy guarantee and may expose IP addresses, object sizes, timing, cache behavior, and routing metadata. Deployments requiring network-layer privacy must add separate mechanisms such as anonymous communication, private retrieval, or traffic-analysis resistance. Deployments must also comply with applicable data protection, accessibility, credential governance, and sector-specific requirements.

\noindent\textbf{Decision to Proceed.} We did not recruit or observe people, use real identities or personal data, or test live services. The study involved no research participants and required no human-subject consent. The evaluation used synthetic credentials and controlled infrastructure. This limits the study's direct risks but does not remove our responsibility to consider how later deployment could affect people. Residual risks include premature deployment, misuse of reduced traceability, and leakage outside the proof transcript.

Although a future use of \textsc{ShadowPath} by the general public may lead to several potential consequences, we argue that the contributions of \textsc{ShadowPath} outweigh the identified risks. We adhered to the principles of \textit{Respect for Persons}, \textit{Justice}, \textit{Beneficence}, and \textit{Respect for Law and Public Interest}. Furthermore, we publish the artifacts of this study to allow independent verification by our research community and potential future use in live services. 
After considering the ethical principles outlined by the Menlo Report, we conclude that conducting and publishing this study is ethically justified.

\endgroup



%
\bibliographystyle{IEEEtranS}
\bibliography{IEEEabrv, refs}

\raggedbottom
\appendices
\section{Generative AI Considerations} In accordance with the NDSS guidelines regarding generative AI, the authors disclose the use of AI software tools during the preparation of this manuscript and its supporting materials. The foundational research, system design, and underlying intellectual contributions of \textsc{ShadowPath} were developed independently during a one-year research project. We used the listed models to support our research for the following use cases: 

\begin{itemize} 
\item \textbf{Claude Opus (v4.8):} To assist with programming (inc. debugging); and artifact packaging and reproducibility checks.
\item \textbf{Gemini and Gemini Pro (v3.1):} For editorial purposes throughout the paper. 
\item \textbf{ChatGPT (v5.6sol):} For editorial purposes throughout the paper and paper organization. In addition, we used these models to help with \LaTeX\ commands (such as tables) and to debug \LaTeX\ errors and warnings. 
\end{itemize}

These tools did not synthesize, fabricate, or select benchmark outcomes. All reported measurements were produced by executing the documented evaluation under the stated measurement protocol. The authors have thoroughly reviewed, edited, and verified all AI-assisted outputs. The resulting work represents our novel intellectual contribution, and the authors take full responsibility for the accuracy and correctness of this paper. 

\section{\textsc{ShadowPath}-Verkle Setup and Initialization}
\label{app:setup_retrieval}
\newcommand{\appformulanote}[1]{%
  \textcolor{ieeeblue}{\triangleright\;\text{\normalfont #1}}%
}

Algorithm~\ref{alg:setup} summarizes the initialization of the evaluated Verkle prototype. It creates the registry and proof-system parameters, publishes the initial complete summary through IPFS, and signs the IPNS binding to its content identifier.

\noindent\textbf{Implementation scope.} Groth16 requires circuit-specific parameters, whereas PLONK supports universal, updatable parameters. The evaluated prototype generates the inner Verkle KZG SRS and both outer proof-system parameter sets locally with operating-system randomness. Consequently, the party running setup knows the resulting toxic-waste secrets. Setup and key derivation are excluded from timed proving and verification. Deployment must replace these development parameters with ceremony-derived sound parameters. The deterministic Rust--Go bridge imports arkworks little-endian encodings and reproduces the issuer commitment and root bit-for-bit.

\begin{algorithm}[H]
\caption{\textsc{ShadowPath}-Verkle Initialization}
\label{alg:setup}
\small
\begin{algorithmic}[1]

\Require Security parameter $\lambda$, public issuer domain $I$, address space $N=k^d$, Verkle branching factor $k$, depth $d$, proof backend $\mathcal P\in\{\mathsf{Groth16},\mathsf{PLONK}\}$
\Ensure Public parameters $\mathsf{pp}$, issuer state $\mathsf{st}_{\mathsf{iss}}$, initial root $R_0$, authoritative IPNS name $P_{\mathsf{ipns}}$

\State $(\mathsf{sk}_I,\mathsf{pk}_I)
\leftarrow\mathsf{Credential.KeyGen}(1^\lambda)$
\Comment{Generate the issuer-signing key pair}

\State $(\mathsf{sk}_{\mathsf{ipns}},
\mathsf{pk}_{\mathsf{ipns}})
\leftarrow\mathsf{IPNS.KeyGen}(1^\lambda)$
\Comment{Generate the state-publication key pair}

\State $\mathsf{grp}\leftarrow
\mathsf{GroupSetup}(1^\lambda)$
\Comment{Initialize the required pairing groups}

\State $\mathsf{srs}_{\mathsf{Verkle}}\leftarrow
\mathsf{KZG.Setup}(1^\lambda,k)$
\Comment{Generate evaluation KZG parameters for arity $k$}

\State $T_0\leftarrow
\mathsf{EmptyVerkleTree}
(N,k,d,\mathsf{srs}_{\mathsf{Verkle}})$
\Comment{Initialize the sparse revocation registry}

\State $R_0\leftarrow\mathsf{RootCommitment}(T_0)$

\State $(\mathsf{pk}_{\mathcal P},
\mathsf{vk}_{\mathcal P})
\leftarrow
\mathcal P.\mathsf{Setup}
\!\left(1^\lambda,\mathcal R_{\mathsf{Verkle}}\right)$
\Comment{Generate keys for the complete status relation}

\State $\mathsf{sum}_0\leftarrow
\mathsf{ExtractSummary}(T_0)$
\Comment{Serialize the materialized registry state}

\State $t_0\leftarrow\mathsf{Clock.Now}()$;
$\mathsf{prev}_0\leftarrow\bot$

\State $\mu_0\leftarrow
(0,t_0,\mathsf{prev}_0,\mathsf{sum}_0)$
\Comment{Construct the initial epoch payload}

\State $\chi_0\leftarrow\mathsf{Checksum}(\mu_0)$

\State $\mathsf{blob}_0\leftarrow
\mathsf{Serialize}(\mu_0,\chi_0)$
\Comment{Construct the versioned epoch object}

\State $\mathsf{cid}_0\leftarrow
\mathsf{IPFS.Publish}(\mathsf{blob}_0)$
\Comment{Publish the content-addressed object}

\State $P_{\mathsf{ipns}}\leftarrow
\mathsf{IPNS.Name}(\mathsf{pk}_{\mathsf{ipns}})$

\State $\mathsf{IPNS.Publish}
(\mathsf{sk}_{\mathsf{ipns}},
 P_{\mathsf{ipns}},
 \mathsf{cid}_0)$
\Comment{Sign and publish the authoritative CID binding}

\State $\mathsf{pp}\leftarrow
\bigl(
 I,N,k,d,
 \mathsf{grp},
 \mathsf{srs}_{\mathsf{Verkle}},
 \mathsf{pk}_I,
 \mathsf{pk}_{\mathsf{ipns}},
 \mathsf{pk}_{\mathcal P},
 \mathsf{vk}_{\mathcal P},
 P_{\mathsf{ipns}}
\bigr)$

\State $\mathsf{st}_{\mathsf{iss}}\leftarrow
\bigl(
 \mathsf{sk}_I,
 \mathsf{sk}_{\mathsf{ipns}},
 T_0,
 0,
 R_0,
 \mathsf{cid}_0
\bigr)$

\State \Return
$(\mathsf{pp},\mathsf{st}_{\mathsf{iss}},R_0,
P_{\mathsf{ipns}})$

\end{algorithmic}
\end{algorithm}

\section{\textsc{ShadowPath}-Verkle Construction Details}
\label{app:verkle-details}

\subsection{Verkle Relation Derivation}
\label{app:mode-b-derivation}

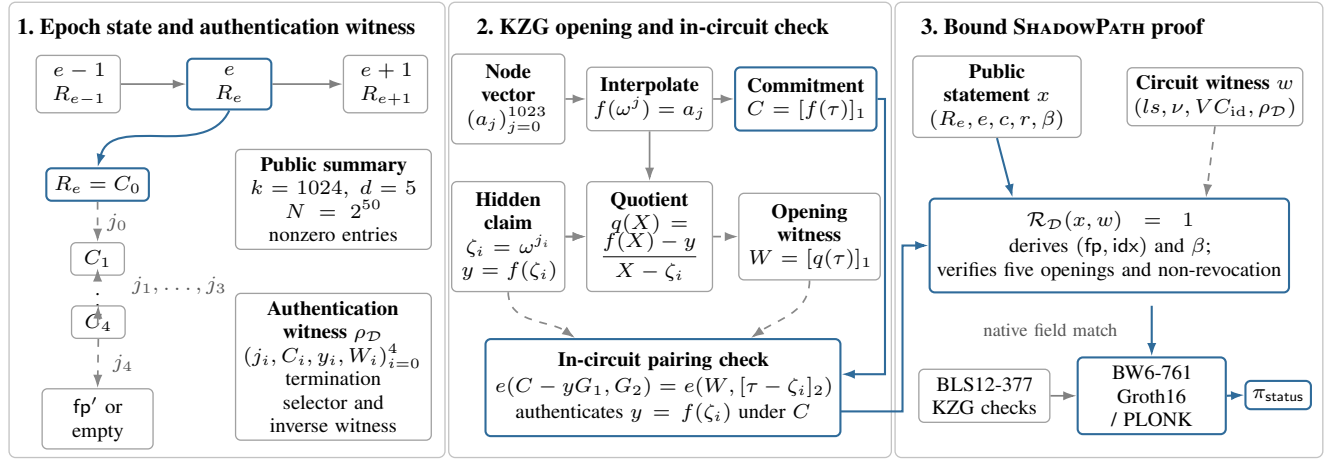
\begin{figure*}[!t]
\centering
\resizebox{\textwidth}{!}{\definecolor{preblue}{HTML}{2F6B9A}
\definecolor{pregray}{HTML}{555555}

\begingroup
\newcommand{\modebfont}{\rmfamily\fontsize{7.75pt}{8.95pt}\selectfont}
\newcommand{\modebtitlefont}{\rmfamily\bfseries\fontsize{8.65pt}{9.85pt}\selectfont}
\newcommand{\modebsmallfont}{\rmfamily\fontsize{7pt}{8.1pt}\selectfont}
\begin{tikzpicture}[
  x=1cm,
  y=1cm,
  font=\modebfont,
  panel/.style={
    draw=pregray!35,
    rounded corners=2.5pt,
    line width=.5pt
  },
  title/.style={
    anchor=north west,
    inner sep=0pt,
    outer sep=0pt,
    font=\modebtitlefont
  },
  box/.style={
    draw=pregray!55,
    rounded corners=2pt,
    line width=.5pt,
    fill=white,
    align=center,
    inner xsep=3pt,
    inner ysep=3pt
  },
  public/.style={box},
  private/.style={box},
  crypto/.style={box},
  accent/.style={
    box,
    draw=preblue,
    line width=.8pt
  },
  arrow/.style={
    -{Latex[length=1.8mm,width=1.25mm]},
    draw=pregray!75,
    line width=.65pt
  },
  publicarrow/.style={
    arrow,
    draw=preblue,
    line width=.8pt
  },
  privatearrow/.style={
    arrow,
    draw=pregray!70,
    dashed
  },
  cryptoarrow/.style={
    arrow,
    draw=pregray!70
  },
  note/.style={
    align=center,
    text=pregray,
    font=\modebfont
  }
]

\path[use as bounding box]
  (-.02,-.24) rectangle (17.67,5.92);


\draw[panel] (-0.01,-0.22) rectangle (5.85,5.9);
\draw[panel] (5.9,-0.22) rectangle (11.85,5.9);
\draw[panel] (11.90,-.22) rectangle (17.65,5.90);


\node[title, minimum width=145pt] at (0.1,5.68)
  {1. Epoch state and authentication witness};

\node[title] at (6.28,5.68)
  {2. KZG opening and in-circuit check};

\node[title] at (12.25,5.68)
  {3. Bound \textsc{ShadowPath} proof};


\node[
  box,
  text width=.92cm,
  minimum height=.55cm
] (em1) at (.92,4.80)
  {$e-1$\\$R_{e-1}$};

\node[
  accent,
  text width=.92cm,
  minimum height=.55cm
] (ee) at (2.95,4.8)
  {$e$\\$R_e$};

\node[
  box,
  text width=.92cm,
  minimum height=.55cm
] (ep1) at (5.05,4.8)
  {$e+1$\\$R_{e+1}$};

\draw[arrow]
  (em1.east) -- (ee.west);

\draw[arrow]
  (ee.east) -- (ep1.west);

\node[
  accent,
  minimum width=1.05cm,
  minimum height=.40cm
] (c0) at (1.18,3.45)
  {$R_e=C_0$};

\node[
  private,
  minimum width=.72cm,
  minimum height=.38cm
] (c1) at (1.18,2.45)
  {$C_1$};

\node (dots) at (1.18,1.94)
  {$\vdots$};

\node[
  private,
  minimum width=.72cm,
  minimum height=.38cm
] (c4) at (1.19,1.6)
  {$C_4$};

\node[
  private,
  text width=1.28cm,
  minimum height=.46cm
] (leaf) at (1.18,0.3)
  {$\mathsf{fp}'$ or empty};

\draw[publicarrow]
  (ee.south)
  to[out=-115,in=75]
  (c0.north);

\draw[privatearrow]
  (c0.south)
  --
  node[
    right,
    note,
    anchor=west
  ] {$j_0$}
  (c1.north);

\draw[privatearrow]
  (c1.south) -- (dots.north);

\node[
  note,
  anchor=west
] at (1.52,2.1)
  {$j_1,\ldots,j_3$};

\draw[privatearrow]
  (dots.south) -- (c4.north);

\draw[privatearrow]
  (c4.south)
  --
  node[
    midway,
    right,
    note,
    anchor=west,
    xshift=2pt
  ] {$j_4$}
  (leaf.north);

\node[
  public,
  text width=2.42cm,
  minimum height=1.10cm,
  font=\modebfont
] (summary) at (4.35,3.25)
  {
    \textbf{Public summary}\\
    $k=1024,\ d=5$\\
    $N=2^{50}$\\
    nonzero entries
  };

\node[
  private,
  text width=2.42cm,
  minimum height=1.15cm,
  font=\modebfont
] (authwitness) at (4.35,1)
  {
    \textbf{Authentication witness $\rho_{\mathcal D}$}\\
    $(j_i,C_i,y_i,W_i)_{i=0}^{4}$\\
    termination selector and\\
    inverse witness
  };


\node[
  box,
  text width=1.38cm,
  minimum height=.82cm,
  inner xsep=2pt,
  inner ysep=4pt,
  font=\modebfont
] (vec) at (6.7,4.6)
  {
    \textbf{Node vector}\\
    $(a_j)_{j=0}^{1023}$
  };

\node[
  crypto,
  text width=1.55cm,
  minimum height=.82cm,
  inner xsep=2pt,
  inner ysep=4pt,
  font=\modebfont
] (poly) at (8.6,4.6)
  {
    \textbf{Interpolate}\\
    $f(\omega^j)=a_j$
  };

\node[
  accent,
  text width=1.76cm,
  minimum height=.82cm,
  inner xsep=2pt,
  inner ysep=4pt,
  font=\modebfont
] (commit) at (10.7,4.6)
  {
    \textbf{Commitment}\\
    $C=[f(\tau)]_1$
  };

\draw[cryptoarrow]
  (vec.east) -- (poly.west);

\draw[cryptoarrow]
  (poly.east) -- (commit.west);

\node[
  private,
  text width=1.38cm,
  minimum height=.98cm,
  inner xsep=2pt,
  inner ysep=4pt,
  font=\modebfont
] (claim) at (6.7,2.75)
  {
    \textbf{Hidden claim}\\
    $\zeta_i=\omega^{j_i}$
    $y=f(\zeta_i)$
  };

\node[
  crypto,
  text width=1.55cm,
  minimum height=.98cm,
  inner xsep=2pt,
  inner ysep=4pt,
  font=\modebfont
] (quotient) at (8.6,2.75)
  {
    \textbf{Quotient}\\
    $q(X)=$\\[-1pt]
    $\dfrac{f(X)-y}{X-\zeta_i}$
  };

\node[
  private,
  text width=1.76cm,
  minimum height=.98cm,
  inner xsep=2pt,
  inner ysep=4pt,
  font=\modebfont
] (opening) at (10.75,2.75)
  {
    \textbf{Opening witness}\\
    $W=[q(\tau)]_1$
  };

\draw[cryptoarrow]
  (poly.south) -- (quotient.north);

\draw[privatearrow]
  (claim.east) -- (quotient.west);
\draw[privatearrow]
  (claim.east) -- (quotient.west);

\draw[privatearrow]
  (quotient.east) -- (opening.west);

\node[
  accent,
  text width=4.56cm,
  minimum height=1.02cm,
  inner ysep=5pt,
  font=\modebfont
] (pairing) at (8.78,.72)
  {
    \textbf{In-circuit pairing check}\\[1pt]
    $e(C-yG_1,G_2)
      =
      e\!\left(W,[\tau-\zeta_i]_2\right)$\\[1pt]
    authenticates $y=f(\zeta_i)$ under $C$
  };

\draw[publicarrow]
  (commit.east)
  -- ++(0.1,0)
  |- ([yshift=.18cm]pairing.east);

\draw[privatearrow]
  (claim.south)
  to[out=-90,in=145]
  ([xshift=-1.18cm]pairing.north);

\draw[privatearrow]
  (opening.south)
  to[out=-90,in=35]
  ([xshift=1.18cm]pairing.north);


\node[
  public,
  text width=2.05cm,
  minimum height=.72cm,
  font=\modebfont
] (pub) at (13.27,4.63)
  {
    \textbf{Public statement $x$}\\
    $(R_e,e,c,r,\beta)$
  };

\node[
  private,
  text width=2.15cm,
  minimum height=.72cm,
  font=\modebfont
] (priv) at (16.20,4.63)
  {
    \textbf{Circuit witness $w$}\\
    $(ls,\nu,VC_{\mathrm{id}},\rho_{\mathcal D})$
  };

\node[
  accent,
  text width=4.62cm,
  minimum height=1.26cm,
  font=\modebfont
] (relation) at (14.78,2.63)
  {
    \textbf{$\mathcal R_{\mathcal D}(x,w)=1$}\\[1pt]
    derives $(\mathsf{fp},\mathsf{idx})$ and $\beta$;\\
    verifies five openings and non-revocation
  };

\draw[publicarrow]
  (pub.south)
  --
  ([xshift=-1.30cm]relation.north);

\draw[privatearrow]
  (priv.south)
  --
  ([xshift=1.30cm]relation.north);

\node[
  crypto,
  text width=1.55cm,
  minimum height=.64cm,
  font=\modebfont
] (inner) at (13.10,.61)
  {
    BLS12-377\\
    KZG checks
  };

\node[
  accent,
  text width=1.78cm,
  minimum height=.64cm,
  font=\modebfont
] (outer) at (15.35,.61)
  {
    BW6-761\\
    Groth16 / PLONK
  };

\node[accent, minimum width=.88cm, minimum height=.38cm, inner xsep=2pt, font=\modebfont] (proof) at (17.05,.62)
  {$\pi_{\mathsf{status}}$};

\draw[publicarrow]
  (11.17,0.4)
  -- ++(0.83,0)
  |- (relation.west);

\draw[publicarrow]
  (15.35,1.82)
  -- (outer.north);

\draw[arrow]
  (inner.east)
  --
  node[above=19pt, note, font=\modebsmallfont, at start] {native field match}
  (outer.west);

\draw[publicarrow]
  (outer.east) -- (proof.west);

\end{tikzpicture}
\endgroup}
\caption{\textsc{ShadowPath}-Verkle relation: private path, in-circuit
KZG check, and bound outer proof.}
\label{fig:shadowpath_relation}
\end{figure*}

Figure~\ref{fig:shadowpath_relation} relates the private authentication witness, the in-circuit KZG checks, and the outer proof used by the evaluated Verkle relation.

The circuit derives

\begin{equation*}
\begin{aligned}
u&=H_{\mathsf{cred}}(ls,I), &
VC_{\mathrm{id}}&=H_{\mathsf{nonce}}(u,\nu),\\
h&=H_{\mathsf{addr}}(VC_{\mathrm{id}}), &
(\mathsf{fp},\mathsf{idx})&=\mathsf{Split}(h),\\
\gamma&=H_{\mathsf{ctx}}(c,e), &
\beta&=H_{\mathsf{ent}}
  (H_{\mathsf{bind}}(VC_{\mathrm{id}},\gamma),r).
\end{aligned}
\end{equation*}

These Poseidon invocations use disjoint leading tags: credential-base, credential-nonce, address, context, binding, entropy, and commitment encoding use $101,\ldots,107$, respectively. Rust--Go conformance tests fix the derived field elements and root bit-for-bit. The circuit decomposes the BW6-761 scalar $h$ into 377 little-endian bits and enforces that the reconstructed integer is below the scalar-field modulus. This reducedness constraint rejects the congruent non-canonical representation $h+q$. Bits $0$--$127$ form $\mathsf{fp}$, bits $128$--$177$ form the five 10-bit digits $(j_0,\ldots,j_4)$, and bits $178$--$376$ are unused by the evaluated 50-bit profile; wider allocations are analyzed below.

For an affine BLS12-377 commitment $C=(X_C,Y_C)$, the circuit encodes the corresponding child commitment as
\begin{equation*}
\begin{gathered}
\begin{aligned}
u_C&=H_{\mathsf{Poseidon}}(D_{\mathsf{enc}},X_C,Y_C),\\
\mathsf{Enc}(C)&=\mathsf{low}_{252}(u_C)
  \in\mathbb F_r(\mathrm{BLS12\text{-}377}).
\end{aligned}\\[-0.2ex]
\appformulanote{commitment encoding}
\end{gathered}
\end{equation*}
Before the private termination level, the circuit recomputes this coordinate hash and constrains $y_i=\mathsf{Enc}(C_{i+1})$. The truncation is canonical in the BLS12-377 scalar field; path binding assumes that this encoding is collision-resistant.

Each KZG opening has three conceptual components: $C$ commits to one node, $(\zeta_i,y)$ claims that the node stores value~$y$ at hidden position~$\zeta_i$, and $W$ authenticates that claim. For one Verkle node, let $(a_j)_{j=0}^{1023}$ be its evaluation vector and let $f(\omega^j)=a_j$. Registry construction commits to $C=[f(\tau)]_1$. For hidden position $\zeta_i=\omega^{j_i}$ and claimed value~$y$, the holder constructs
\begin{equation*}
\begin{gathered}
q(X)=\frac{f(X)-y}{X-\zeta_i},
\qquad
W=[q(\tau)]_1.\\[-0.2ex]
\appformulanote{KZG quotient witness}
\end{gathered}
\end{equation*}
If $y=f(\zeta_i)$, the numerator is divisible by $X-\zeta_i$, and $W$ commits to the resulting quotient. The Verkle circuit binds $(\zeta_i,y)$ to the private path and verifies the resulting opening equation.

In Equation~\eqref{eq:mode-b-kzg}, the first line is the KZG opening check, while the second constructs its verifier-side $G_2$ term:

\begin{equation}
\label{eq:mode-b-kzg}
\begin{gathered}
\begin{aligned}
e(C-yG_1,G_2)
  &=e\!\left(W,[\tau-\zeta_i]_2\right),\\[-0.2ex]
[\tau-\zeta_i]_2&=[\tau]_2-\zeta_i G_2.
\end{aligned}\\[-0.2ex]
\appformulanote{KZG opening verification}
\end{gathered}
\end{equation}
Under KZG binding, Equation~(1) authenticates $y$ as the value
committed at~$\zeta_i$. The implementation evaluates this identity over BLS12-377 inside a BW6-761 outer circuit, using the two-chain equality
\begin{equation*}
\begin{gathered}
\mathbb F_q(\mathrm{BLS12\text{-}377})
=\mathbb F_r(\mathrm{BW6\text{-}761}).\\[-0.2ex]
\appformulanote{two-chain field match}
\end{gathered}
\end{equation*}
The field match removes generic foreign-base-field emulation but not the five pairing-based opening checks.

With uniformly sampled 50-bit candidate indices, the birthday estimate for at least one collision among \(n\) issuance attempts before rejection sampling is
\begin{equation*}
\begin{gathered}
p_{\mathsf{coll}}(n)\approx
1-\exp\!\left(-\frac{n(n-1)}{2^{51}}\right).\\[-0.2ex]
\appformulanote{50-bit collision estimate}
\end{gathered}
\end{equation*}
It yields $4.44\times10^{-6}$ at $10^5$, $4.44\times10^{-4}$ at $10^6$, and $4.34\times10^{-2}$ at $10^7$. Rejection sampling gives next-issuance retry probability $n/2^{50}$, which is $8.88\times10^{-10}$ at $n=10^6$. With $k=1024$, 64- and 128-bit indices require depths $d=7$ and $d=13$. The circuit must constrain the six unused bits of the 70-bit path and the two unused bits of the 130-bit path to zero. At $n=10^8$, their birthday estimates are $2.71\times10^{-4}$ and $1.47\times10^{-23}$. Relative to $d=5$, they add two and eight KZG openings (40\% and 160\% more commitments, proofs, points, and subgroup checks), whereas root-link checks increase from four to six and twelve. Proof size remains backend-constant per compiled profile, but circuits, keys, complete circuit-witness size, and proving work grow. The observed PLONK domain transitions preclude linear extrapolation of mobile latency. Alternatively, retaining five levels for at least 64 bits requires $k\geq2^{13}$, thus requiring eight-times-wider node vectors and SRS.

The lifecycle store reserves indices across issuances and retries an occupied index with a fresh authenticated credential nonce. We scope the evaluated 50-bit profile to at most $10^6$ reserved indices per issuer domain as an operating-profile boundary, not as a correctness limit from collisions: rejection sampling preserves uniqueness, and the next-issuance retry probability is still $8.88\times10^{-10}$ at that boundary. We make no performance or anonymity-set claim beyond it. Larger deployments can use a wider profile, a committed collision bucket, authenticated re-salting, or multiple-choice hashing; these extensions are not implemented.

\subsection{Sparse Summary and Address Mapping}
\label{app:verkle-summary}

\begin{figure}[!ht]
\centering
\definecolor{spblue}{HTML}{56B4E9}
\definecolor{spgreen}{HTML}{009E73}
\definecolor{spgray}{HTML}{555555}

\begingroup
\newcommand{\summaryfont}{%
  \rmfamily\fontsize{9.2pt}{10.4pt}\selectfont}
\newcommand{\summaryboldfont}{%
  \rmfamily\bfseries\fontsize{10.2pt}{11.4pt}\selectfont}

\noindent\resizebox{\columnwidth}{!}{%
\begin{tikzpicture}[
  font=\summaryfont,
  commitment/.style={
    circle,
    draw=spgray,
    line width=.7pt,
    fill=spblue!10,
    minimum size=9mm,
    inner sep=1pt
  },
  omitted/.style={
    circle,
    draw=spgray,
    dashed,
    line width=.6pt,
    minimum size=7mm,
    inner sep=1pt,
    text=spgray
  },
  leaf/.style={
    circle,
    draw=spgreen,
    line width=.8pt,
    fill=spgreen!10,
    minimum size=8.5mm,
    inner sep=1pt
  },
  record/.style={
    draw=spgray,
    rounded corners=1.5pt,
    line width=.6pt,
    fill=black!2,
    minimum height=7mm,
    align=center,
    inner xsep=3pt,
    inner ysep=2.5pt
  },
  stored/.style={
    draw=spgreen,
    line width=1.15pt
  },
  absent/.style={
    draw=spgray,
    dashed,
    line width=.65pt
  },
  mapping/.style={
    -{Latex[length=1.8mm]},
    draw=spblue,
    dashed,
    line width=.8pt
  }
]
\path[use as bounding box]
  (-1.32,-2.74) rectangle (9.81,4.78);

\node[font=\summaryboldfont,anchor=west]
  at (-.5,4.55) {Sparse Verkle registry};

\node[commitment] (root) at (2.0,3.85) {$R_e$};

\node[commitment] (c1) at (.8,2.55) {$C_1$};
\node[omitted]    (e1) at (2.0,2.55) {$\emptyset$};
\node[commitment] (c3) at (3.2,2.55) {$C_3$};

\node[omitted] (e2) at (.15,1.25) {$\emptyset$};
\node[leaf]    (va) at (1.15,1.25) {$v_a$};
\node[leaf]    (vb) at (2.45,1.25) {$v_b$};
\node[omitted] (e3) at (3.45,1.4) {$\emptyset$};

\foreach \a/\b in {root/c1,root/c3,c1/va,c3/vb}
  \draw[stored] (\a)--(\b);

\foreach \a/\b in {root/e1,c1/e2,c3/e3}
  \draw[absent] (\a)--(\b);

\node[anchor=east,text=spgray] at (-0.2,3.85) {level 0};
\node[anchor=east,text=spgray] at (-0.2,2.55) {level 1};
\node[anchor=east,text=spgray] at (-0.2,1.25) {leaves};


\node[font=\summaryboldfont,anchor=west]
  at (4.3,4.55) {Published complete summary};

\node[record] (r0) at (6.05,3.75)
  {$[\,],\,R_e,\,\{1{:}\mathsf{Enc}(C_1),\,3{:}\mathsf{Enc}(C_3)\}$};

\node[record,minimum width=27mm] (r1) at (5.9,2.65)
  {$[1],\,C_1,\,\{2{:}v_a\}$};

\node[record,minimum width=27mm] (r3) at (5.9,1.55)
  {$[3],\,C_3,\,\{0{:}v_b\}$};

\node[record,minimum width=20mm] (rla) at (5.15,.45)
  {$[1,2]{:}v_a$};

\node[record,minimum width=20mm] (rlb) at (6.75,.45)
  {$[3,0]{:}v_b$};

\draw[mapping] (root) to[out=15,in=180] (r0);
\draw[mapping] (c1)   to[out=35, in=162]  (r1);
\draw[mapping] (c3)   to[out=-5,in=180] (r3);
\draw[mapping] (va)   to[out=-18,in=180] (rla);
\draw[mapping] (vb)   to[out=-22,in=155] (rlb);


\draw[spgreen!75!black, rounded corners=2pt, line width=.7pt, fill=spgreen!5] (-0.45,0) rectangle (8.89,-2.61);

\node[anchor=west,font=\summaryboldfont]
  at (-0.43,-0.25) {Legend};

\node[anchor=west,font=\summaryfont]
  at (-0.27,-0.6) {\textit{Registry symbols}};

\node[anchor=west,font=\summaryfont]
  at (3.45,-0.58) {\textit{Summary notation}};

\node[anchor=west] at (-0.37,-0.95)
  {$R_e$ \quad epoch root};

\node[anchor=west] at (-0.37,-1.3)
  {$C_i$ \quad internal commitment};

\node[anchor=west] at (-0.3,-1.65)
  {$v_i$ \quad revoked-leaf value};

\node[anchor=west] at (-0.29,-2.05)
  {$\emptyset$ \quad absent/zero child};

\node[anchor=west] at (3.35,-0.93)
  {$[i,\ldots]$ \quad tree path};

\node[anchor=west] at (3.35,-1.35)
  {$\{j{:}x\}$ \quad stored child entry};

\node[anchor=west] at (3.4,-1.75)
  {$\mathsf{Enc}(C)$ \quad child-commitment encoding};

\fill[spgreen!75!black] (-0.08,-2.4) circle[radius=1.15pt];

\node[anchor=west]
  at (0.05,-2.4) {Stored leaf $=$ revoked};

\fill[spgray] (3.68,-2.15) circle[radius=1.15pt];

\node[anchor=west]
  at (3.8,-2.15) {Omitted child $=$ non-revoked};
\end{tikzpicture}%
}
\endgroup
\caption{Sparse encoding of the complete Verkle summary. Stored leaves are
revoked; omitted children are non-revoked.}
\label{fig:sp_verkle_summary}
\end{figure}
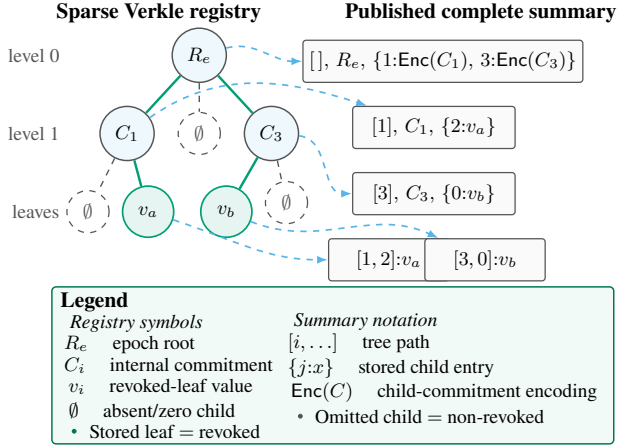

To support the holder-side non-revocation proof described in Section~\ref{sec:implementation}, the issuer distributes a compressed representation of the revocation state. Figure~\ref{fig:sp_verkle_summary} illustrates this sparse encoding.

\begin{figure}[!ht]
\centering
\definecolor{spblue}{HTML}{56B4E9}
\definecolor{spgreen}{HTML}{009E73}
\definecolor{spgray}{HTML}{555555}

\begin{tikzpicture}[
  font=\small,
  line cap=round,
  line join=round,
  rowlabel/.style={
    anchor=west,
    text=spgray
  },
  digit/.style={
    draw=spblue!70!black,
    fill=spblue!8,
    rounded corners=1pt,
    line width=.55pt,
    minimum width=7mm,
    minimum height=6mm,
    align=center
  },
  opening/.style={
    draw=spgray!70,
    fill=spgray!4,
    rounded corners=1pt,
    line width=.5pt,
    minimum width=3.45cm,
    minimum height=6mm,
    align=center
  },
  result/.style={
    draw=spgray!70,
    fill=white,
    rounded corners=1pt,
    line width=.5pt,
    minimum width=2.15cm,
    minimum height=6mm,
    align=center
  },
  terminal/.style={
    result,
    draw=spgreen!75!black,
    fill=spgreen!8
  },
  condition/.style={
    draw=spgreen!75!black,
    fill=spgreen!5,
    rounded corners=1.5pt,
    line width=.55pt,
    minimum width=3.75cm,
    minimum height=7mm,
    align=center
  },
  arrow/.style={
    -{Latex[length=1.5mm,width=1mm]},
    draw=spgray!75,
    line width=.55pt
  },
  selectarrow/.style={
    -{Latex[length=1.5mm,width=1mm]},
    draw=spblue!70!black,
    line width=.7pt
  },
  continuation/.style={
    -{Latex[length=1.4mm,width=.95mm]},
    draw=spgray!60,
    densely dashed,
    line width=.5pt
  }
]

\path[use as bounding box]
  (-0.45,-4.59) rectangle (8.44,0.63);

\node[anchor=west] at (-0.1,0.3)
  {$\mathsf{idx}\mapsto(j_0,j_1,j_2,j_3,j_4)$};

\node[anchor=east,text=spgray] at (7.97,0.3)
  {$j_i\in\{0,\ldots,1023\}$};

\node[rowlabel] at (-0.45,-0.47)
  {level 0: $R_e$};

\node[digit] (j0) at (1.57,-0.44)
  {$j_0$};

\node[opening] (o0) at (4.1,-0.44)
  {$y_0=f_0(\omega^{j_0})$};

\node[result] (r0) at (7.35,-0.44)
  {$\mathsf{Enc}(C_1)$ or $0$};

\draw[selectarrow] (j0.east)--(o0.west);
\draw[arrow] (o0.east)--(r0.west);

\node[rowlabel] at (-0.45,-1.5)
  {level 1: $C_1$};

\node[digit] (j1) at (1.57,-1.5)
  {$j_1$};

\node[opening] (o1) at (4.1,-1.5)
  {$y_1=f_1(\omega^{j_1})$};

\node[result] (r1) at (7.35,-1.5)
  {$\mathsf{Enc}(C_2)$ or $0$};

\draw[selectarrow] (j1.east)--(o1.west);
\draw[arrow] (o1.east)--(r1.west);
\draw[continuation] (r0.south)--(r1.north);

\node[text=spgray] at (4.1,-2.2)
  {$\vdots$};

\node[text=spgray] at (7.35,-2.2)
  {$\vdots$};

\node[rowlabel] at (-0.45,-2.92)
  {level 4: $C_4$};

\node[digit] (j4) at (1.57,-2.92)
  {$j_4$};

\node[opening] (o4) at (4.1,-2.92)
  {$y_4=f_4(\omega^{j_4})$};

\node[terminal] (r4) at (7.35,-2.92)
  {$\mathsf{fp}'$ or $0$};

\draw[selectarrow] (j4.east)--(o4.west);
\draw[arrow] (o4.east)--(r4.west);

\node[condition, minimum width=131pt] at (2,-3.95)
  {authenticated $0$ before the leaf\\
   $\Rightarrow$ early non-revocation};

\node[condition, minimum width=112pt] at (6.43,-3.94)
  {terminal $\mathsf{fp}'\neq\mathsf{fp}$\\
   $\Rightarrow$ non-revocation};

\end{tikzpicture}
\caption{Verkle path authentication for the evaluated $k=1024,d=5$
profile. Each digit selects one child, and the circuit verifies one KZG
opening per level. An authenticated zero child establishes early
non-revocation; subsequent levels use valid dummy openings. Otherwise, a
terminal fingerprint mismatch establishes non-revocation.}
\label{fig:digit_mapping}
\end{figure}
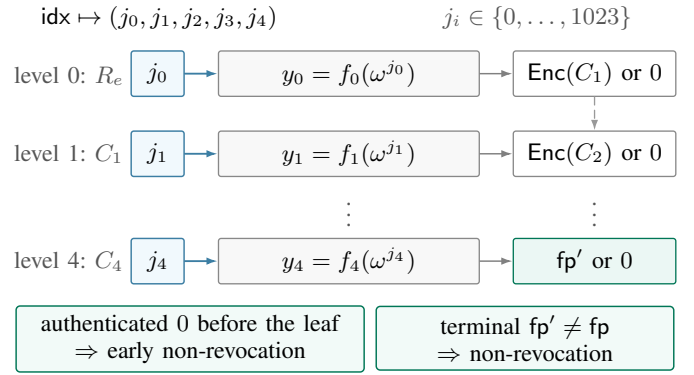

The complete summary used for bootstrap or recovery contains materialized nodes and revoked leaves. Omitting empty subtrees reduces its size to $\mathcal{O}(N_{\mathrm{rev}})$ for fixed depth~$d$; this compression does not make state retrieval anonymous. Figure~\ref{fig:sp_bit_allocation} shows how the circuit decomposes the canonical Poseidon output into a 128-bit fingerprint $\mathsf{fp}$ and a 50-bit registry index~$\mathsf{idx}$. Figure~\ref{fig:digit_mapping} then connects this index to the registry structure: each 10-bit digit~$j_i$ of $\mathsf{idx}$ selects one of the $k=1024$ children at level~$i$, so the five digits define exactly one root-to-leaf route.

\begin{figure}[!ht]
\centering
\begingroup
\newcommand{\bitallocationfont}{\rmfamily\fontsize{8.2pt}{9.4pt}\selectfont}
\begin{tikzpicture}[
  font=\bitallocationfont,
  line cap=round,
  line join=round,
  box/.style={draw=gray!70,line width=.45pt,minimum height=7mm,
              align=center,inner sep=1.5pt},
  guide/.style={draw=blue!55!gray,line width=.55pt}
]
\path[use as bounding box] (-.05,-2.02) rectangle (8.05,1.36);

\node[anchor=west] at (0,1.18)
  {canonical $h\in\mathbb F_r(\mathrm{BW6\text{-}761})$};

\node[box,fill=green!8,minimum width=2.72cm,anchor=west]
  (fp) at (0,.62) {$\mathsf{fp}$};
\node[box,fill=blue!7,minimum width=1.06cm,anchor=west]
  (idx) at (2.72,.62) {$\mathsf{idx}$};
\node[box,fill=gray!7,minimum width=4.22cm,anchor=west]
  (unused) at (3.78,.62) {unused by evaluated profile};

\node[anchor=north] at (1.36,.26) {bits $0$--$127$};
\node[anchor=north] at (3.27,0.26) {bits};
\node[anchor=north] at (3.07,-0.08) {$128$--$177$};
\node[anchor=north] at (5.89,.26) {bits $178$--$376$};

\node[anchor=west] at (0,-1.04) {path digits:};
\foreach \i/\x/\range in {
  0/2.41/{128--137},
  1/3.62/{138--147},
  2/4.83/{148--157},
  3/6.04/{158--167},
  4/7.25/{168--177}}
{
  \node[box,fill=blue!7,minimum width=1.18cm]
    (j\i) at (\x,-1.04) {$j_{\i}$};
  \node[anchor=north,font=\scriptsize] at (\x,-1.40) {\range};
}

\draw[guide] (idx.south east)--(3.78,-.57);
\draw[guide] (1.82,-.57)--(7.84,-.57);
\foreach \x in {2.41,3.62,4.83,6.04,7.25}
  \draw[guide] (\x,-.57)--(\x,-.69);

\node[anchor=west,text=gray!75] at (0,-1.88)
  {little-endian decomposition; $j_0$ selects the root child};
\end{tikzpicture}
\endgroup
\caption{Poseidon allocation: 128-bit fingerprint, five 10-bit path digits,
and unused high bits.}
\label{fig:sp_bit_allocation}
\end{figure}
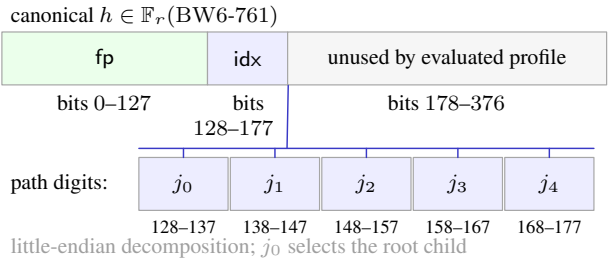

\subsection{Point Validation and Backend Configuration}
\label{app:mode-b-implementation}

The commitments and quotient witnesses form the authentication witness~$\rho_{\mathcal D}$ and therefore the circuit witness~$w$. Before each pairing check, the Verkle circuit enforces BLS12-377 group membership for every commitment and quotient point. Integration tests reject both off-curve and on-curve non-subgroup opening witnesses. SRS import separately checks canonical coordinates, curve and prime-order subgroup membership, the $\tau$-power structure, and the cross-language conformance commitment. These checks enforce the point-validity assumptions used by the standalone status argument in Section~\ref{sec:implementation}.

The authenticated relation uses the BLS12-377$\rightarrow$BW6-761 curve chain with either Groth16 or PLONK.

The EdDSA composition proof of concept pins the issuer public key and verifies an issuer signature with gnark v0.14 over \texttt{twistededwards.BW6\_761} using \texttt{MiMC\_BW6\_761}. The signer encodes $VC_{\mathrm{id}}$ as a canonical fixed-width big-endian scalar. The circuit verifies the signature and the complete Verkle relation over the same private $VC_{\mathrm{id}}$. Its witness contains the issuer signature and status witness~$w$, including authentication witness~$\rho_{\mathcal D}$. Tests reject identifiers authenticated by one check, mismatched issuer keys, and tampered signatures. The construction enforces hidden-identifier equality but is not a production credential-wallet integration.

\section{Supplemental Mobile Results}\label{app:mobile-results}

Table~\ref{tab:mobile_device_matrix} reports the complete physical-device matrix underlying the primary mobile-device results in Section~\ref{sec:results-mobile}. 

\begingroup
\renewcommand{\arraystretch}{1.20}
\begin{table*}[!t]
\centering
\caption{\textsc{ShadowPath}-Verkle results on mobile devices.}
\label{tab:mobile_device_matrix}
\scriptsize
\setlength{\tabcolsep}{4.4pt}
\begin{tabular}{@{}llllrrrr@{}}
\toprule
Device & Order & Backend & {Proving median [range] (\unit{\second})} & {Verification (\unit{\milli\second})} & {Proof (\unit{\byte})} & {OS memory (\unit{\mega\byte})} & {Max OS thermal label} \\
\midrule
SM-A546B (Galaxy A54 5G) & P$\rightarrow$G & PLONK & 125.05 [124.37--128.31] & 39.88 & 1352 & 3627 & light \\
 & P$\rightarrow$G & Groth16 & 10.46 [9.88--10.57] & 29.48 & 484 & 3624 & none \\
 & G$\rightarrow$P & Groth16 & 10.41 [9.88--10.45] & 27.60 & 484 & 3417 & none \\
 & G$\rightarrow$P & PLONK & 124.00 [117.76--174.74] & 47.38 & 1352 & 3675 & none \\
SM-S931B (Galaxy S25) & P$\rightarrow$G & PLONK & 49.03 [43.73--55.89] & 14.57 & 1352 & 3664 & moderate \\
 & P$\rightarrow$G & Groth16 & 2.95 [2.92--3.36] & 25.99 & 484 & 3519 & moderate \\
 & G$\rightarrow$P & Groth16 & 2.90 [2.88--3.06] & 17.87 & 484 & 3009 & moderate \\
 & G$\rightarrow$P & PLONK & 58.04 [46.27--63.98] & 18.63 & 1352 & 3673 & severe \\
SM-S938B (Galaxy S25 Ultra) & P$\rightarrow$G & PLONK & 44.45 [39.80--48.17] & 20.61 & 1352 & 4999 & light \\
 & P$\rightarrow$G & Groth16 & 3.13 [3.02--3.15] & 19.66 & 484 & 4890 & light \\
 & G$\rightarrow$P & Groth16 & 3.07 [3.04--3.16] & 23.99 & 484 & 4345 & none \\
 & G$\rightarrow$P & PLONK & 43.94 [38.23--49.42] & 18.97 & 1352 & 5040 & light \\
iPhone13,2 (iPhone 12) & P$\rightarrow$G & PLONK & 75.73 [72.42--85.80] & 20.62 & 1352 & 2196 & serious \\
 & P$\rightarrow$G & Groth16 & 5.37 [5.26--7.24] & 9.80 & 484 & 486 & serious \\
 & G$\rightarrow$P & Groth16 & 5.33 [5.28--5.83] & 8.85 & 484 & 490 & fair \\
 & G$\rightarrow$P & PLONK & 83.67 [73.90--95.56] & 28.28 & 1352 & 2187 & serious \\
iPhone14,2 (iPhone 13 Pro) & P$\rightarrow$G & PLONK & 56.04 [55.06--67.84] & 15.78 & 1352 & 3447 & serious \\
 & P$\rightarrow$G & Groth16 & 4.58 [4.57--5.95] & 8.53 & 484 & 469 & serious \\
 & G$\rightarrow$P & Groth16 & 4.61 [4.58--5.66] & 8.61 & 484 & 491 & nominal \\
 & G$\rightarrow$P & PLONK & 65.10 [55.40--73.60] & 22.32 & 1352 & 3589 & serious \\
iPhone15,3 (iPhone 14 Pro Max) & P$\rightarrow$G & PLONK & 51.30 [49.00--51.76] & 15.21 & 1352 & 3567 & fair \\
 & P$\rightarrow$G & Groth16 & 4.06 [4.05--4.13] & 7.21 & 484 & 474 & nominal \\
 & G$\rightarrow$P & Groth16 & 4.07 [4.02--4.09] & 6.98 & 484 & 487 & nominal \\
 & G$\rightarrow$P & PLONK & 51.28 [49.01--54.35] & 15.28 & 1352 & 3549 & fair \\
iPhone15,5 (iPhone 15 Plus) & P$\rightarrow$G & PLONK & 52.54 [52.49--56.98] & 12.55 & 1352 & 3516 & nominal \\
 & P$\rightarrow$G & Groth16 & 4.16 [4.10--5.10] & 7.34 & 484 & 481 & nominal \\
 & G$\rightarrow$P & Groth16 & 4.13 [4.10--4.41] & 7.46 & 484 & 489 & nominal \\
 & G$\rightarrow$P & PLONK & 53.17 [52.12--54.04] & 12.39 & 1352 & 3473 & fair \\
iPhone16,2 (iPhone 15 Pro Max) & P$\rightarrow$G & PLONK & 49.68 [48.88--68.17] & 11.09 & 1352 & 3588 & serious \\
 & P$\rightarrow$G & Groth16 & 4.10 [4.02--4.71] & 7.19 & 484 & 467 & serious \\
 & G$\rightarrow$P & Groth16 & 4.28 [4.20--4.47] & 7.72 & 484 & 472 & nominal \\
 & G$\rightarrow$P & PLONK & 51.26 [50.38--60.71] & 12.31 & 1352 & 3597 & serious \\
iPhone18,1 (iPhone 17 Pro) & P$\rightarrow$G & PLONK & 38.85 [38.43--42.19] & 8.15 & 1352 & 3590 & nominal \\
 & P$\rightarrow$G & Groth16 & 3.10 [3.05--3.19] & 5.35 & 484 & 462 & nominal \\
 & G$\rightarrow$P & Groth16 & 3.12 [3.06--3.23] & 5.27 & 484 & 486 & nominal \\
 & G$\rightarrow$P & PLONK & 39.51 [38.57--39.68] & 8.01 & 1352 & 3604 & nominal \\
iPhone18,3 (iPhone 17) & P$\rightarrow$G & PLONK & 44.90 [40.89--46.75] & 9.35 & 1352 & 3599 & fair \\
 & P$\rightarrow$G & Groth16 & 3.13 [3.11--3.23] & 5.50 & 484 & 502 & fair \\
 & G$\rightarrow$P & Groth16 & 3.17 [3.11--3.40] & 5.33 & 484 & 521 & nominal \\
 & G$\rightarrow$P & PLONK & 41.63 [40.29--42.97] & 8.70 & 1352 & 3612 & fair \\
\bottomrule
\end{tabular}
\parbox{0.96\textwidth}{\vspace{1.5mm}\footnotesize Proof-generation time is reported as the median [observed range] over five trials for each backend order. P$\rightarrow$G denotes PLONK followed by Groth16; G$\rightarrow$P denotes Groth16 followed by PLONK. Memory denotes iOS application footprint or Android process RSS and is comparable within each platform, not across platforms; thermal labels are operating-system telemetry.}
\parbox{0.97\textwidth}{\footnotesize\raggedright \emph{Provenance.} Each row contains five retained trials. The importer rejects legacy circuits, simulated devices, incomplete counterbalancing, unexpected proof sizes, duplicate exports, cancellations, and background suspensions. Source identifiers and SHA-256 digests appear in \texttt{result\_e11\_mobile.csv}; Memory and thermal values are operating-system telemetry.}
\end{table*}

\endgroup

\begin{table*}[!t]
\centering
\caption{Expanded notation and data representations for the \textsc{ShadowPath} algorithms.}
\label{tab:algorithm-environment}
\footnotesize
\setlength{\tabcolsep}{5pt}
\renewcommand{\arraystretch}{1.06}

\begin{tabularx}{\textwidth}{@{}
>{\raggedright\arraybackslash}p{0.19\textwidth}
>{\raggedright\arraybackslash}p{0.27\textwidth}
>{\raggedright\arraybackslash}X@{}}
\toprule
\textbf{Object / symbol} &
\textbf{Representation / type} &
\textbf{Role and use} \\
\midrule

\multicolumn{3}{@{}l}{\textit{Common protocol objects}}\\[-0.3ex]

$e$ &
Non-negative integer &
Epoch identifier selected by the verifier and bound into the status proof. \\

$R_e$ &
Backend-specific authenticated root &
Public root for epoch $e$; binds the registry state used by the status relation. \\

$c$ &
Challenge / field element &
Fresh verifier challenge used for session binding and replay protection. \\

$r$ &
Random nonzero field element &
Fresh holder randomizer sampled for each presentation. \\

$\beta$ &
Field element &
Public session-binding value derived from $VC_{\mathrm{id}}$, $c$, $e$, and $r$. \\

$I$ &
Issuer-domain identifier &
Public issuer-domain value used when deriving the hidden credential identifier. \\

$ls$ &
Secret field element &
Holder link secret used to derive $VC_{\mathrm{id}}$. \\

$\nu$ &
Issuer-authenticated nonce &
Credential nonce used with $ls$ and $I$ to derive $VC_{\mathrm{id}}$. \\

$VC_{\mathrm{id}}$ &
Private field element &
Hidden credential identifier shared by credential authentication and status verification. \\

$fp$ &
128-bit integer &
Credential fingerprint derived from $H_{\mathrm{addr}}(VC_{\mathrm{id}})$. \\

$idx$ &
50-bit integer &
Private registry index derived from $H_{\mathrm{addr}}(VC_{\mathrm{id}})$. \\

$D$ &
Registry backend, $\{\mathrm{SMT},\mathrm{Verkle}\}$ &
Selects the authenticated-registry backend. \\

$\rho_D$ &
Backend-specific authentication witness &
Private authentication data used to establish non-revocation under $R_e$. \\

$x$ &
Tuple $(R_e,e,c,r,\beta)$ &
Public statement supplied to the zero-knowledge proof system. \\

$w$ &
Tuple $(ls,\nu,VC_{\mathrm{id}},\rho_D)$ &
Complete private circuit witness. \\

$\pi_{\mathrm{status}}$ &
Zero-knowledge proof &
Verifier-facing proof of authenticated non-revocation and session binding. \\

$\bot$ &
Failure value &
Returned when an algorithm rejects malformed, stale, inconsistent, or otherwise invalid input. \\

\midrule
\multicolumn{3}{@{}l}{\textit{Algorithm 1: State retrieval and local witness reconstruction}}\\[-0.3ex]

$P_{\mathrm{ipns}}$ &
IPNS name / public identifier &
Pinned issuer-controlled name used to resolve the authenticated current epoch object. \\

$st_H$ &
Holder freshness state &
Locally retained state used to detect rollback and enforce freshness policy. \\

$F$ &
Freshness-policy predicate &
Evaluates epoch number, timestamp, predecessor metadata, and root against $st_H$. \\

$cid_e$ &
IPFS content identifier &
Content-addressed identifier of the epoch object resolved through IPNS. \\

$obj_e$ &
Byte string &
Serialized epoch object fetched from IPFS. \\

$payload_e$ &
Epoch payload tuple &
Deserialized epoch payload extracted from $obj_e$. \\

$checksum_e$ &
Digest / checksum &
Integrity value checked before accepting the epoch payload. \\

$t_e$ &
Timestamp &
Publication time associated with epoch $e$. \\

$prev_e$ &
Epoch-predecessor reference &
Reference to the preceding epoch used for freshness and rollback checks. \\

$sum_e$ &
Complete registry summary &
Credential-independent serialized state used for bootstrap or recovery. \\

$state_e$ &
Parsed registry state &
Local registry representation reconstructed from $sum_e$. \\

$\rho_{\mathrm{Verkle}}$ &
Verkle authentication witness &
Credential-specific witness reconstructed locally from $state_e$ and $idx$. \\

\midrule
\multicolumn{3}{@{}l}{\textit{Algorithm 2: Non-revocation proof generation}}\\[-0.3ex]

$P$ &
Proof backend, $\{\mathrm{Groth16},\mathrm{PLONK}\}$ &
Selects the outer proof system; the matched SMT evaluation uses Groth16. \\

$pk_P$ &
Proving key &
Circuit-specific proving material for proof backend $P$. \\

$(fp,idx)$ &
Tuple $(128\text{-bit},50\text{-bit})$ &
Private lookup tuple derived inside the proof from $VC_{\mathrm{id}}$. \\

\midrule
\multicolumn{3}{@{}l}{\textit{Algorithm 3: Verkle setup and initialization}}\\[-0.3ex]

$\lambda$ &
Security parameter &
Controls cryptographic parameter generation. \\

$N$ &
Positive integer, $N=k^d$ &
Registry address-space size. The evaluated profile uses $N=2^{50}$. \\

$k$ &
Positive integer &
Verkle branching factor; $k=1024$ in the evaluated profile. \\

$d$ &
Positive integer &
Verkle depth; $d=5$ in the evaluated profile. \\

$(sk_I,pk_I)$ &
Credential-signing key pair &
Issuer key pair used by the composition proof of concept. \\

$(sk_{\mathrm{ipns}},pk_{\mathrm{ipns}})$ &
IPNS signing key pair &
Authenticates the mutable binding from the issuer IPNS name to the current CID. \\

$grp$ &
Pairing-group parameters &
Public group parameters required by the BLS12-377/BW6-761 construction. \\

$srs_{\mathrm{Verkle}}$ &
KZG structured reference string &
Public setup parameters used for Verkle node commitments and openings. \\

$T_0$ &
Sparse Verkle tree &
Initial empty authenticated revocation registry. \\

$R_0$ &
Verkle root commitment &
Root commitment of the initial empty registry. \\

$(pk_P,vk_P)$ &
Proof-system key pair &
Proving and verification keys for the complete status relation. \\

$sum_0$ &
Complete registry summary &
Serialized initial sparse registry state. \\

$\mu_0$ &
Epoch payload tuple &
Initial epoch metadata and summary before serialization. \\

$\chi_0$ &
Checksum / digest &
Integrity value associated with $\mu_0$. \\

$blob_0$ &
Byte string &
Serialized initial epoch object published through IPFS. \\

$cid_0$ &
IPFS content identifier &
Content identifier returned for $blob_0$. \\

$pp$ &
Public-parameter tuple &
Public system parameters distributed to protocol participants. \\

$st_{\mathrm{iss}}$ &
Issuer-state tuple &
Private issuer state containing signing material, registry state, and current epoch metadata. \\

\bottomrule
\end{tabularx}
\end{table*}

\nobalance
\clearpage
\twocolumn
\section{Artifact Appendix}\label{app:artifact}
This appendix describes how reviewers obtain, configure, and evaluate the artifact. The retained relation suite contains five accepting and thirteen rejecting cases.

\subsection{Description \& Requirements}
The anonymous artifact~\footnote{ShadowPath artifact: \url{https://anonymous.4open.science/r/shadowpath-5097}, Accessed: 2026-08-16} contains the Rust issuer and registry, the self-contained Go/gnark \textsc{ShadowPath}-Verkle prover and verifier, Docker deployment files, the controlled SMT relation, the evaluation harness, and raw provenance. The following requirements define the supported evaluation environment.

\subsubsection{How to access}
The anonymous repository URL is supplied through the submission system. The artifact root contains \texttt{README.md}, \texttt{LICENSE}, \texttt{MANIFEST.sha256}, and the single entry point \texttt{artifact.sh}. If the Available badge is awarded, the AEC-approved version will be deposited in DOI-backed permanent storage and that DOI will replace this review-access statement in the camera-ready appendix.

\subsubsection{Hardware dependencies}
Desktop reproduction requires an x86-64 or ARM64 Linux or macOS host. We recommend 16 GB of RAM and 5 GB of free disk space. Physical-device reruns require one of the named iOS or Android devices and are outside the requested desktop Reproduced scope.

\subsubsection{Software dependencies}
The artifact uses Bash, Git, \texttt{shasum} or \texttt{sha256sum}, Go~1.25 or newer, Rust~1.89 or newer, and Python~3.10 or newer. Network access is required for the initial dependency build. Docker with Compose is needed for IPFS/IPNS and controlled-network experiments. Platform-specific mobile builds additionally require Xcode~16 and XcodeGen or the Android SDK and NDK.

\subsubsection{Benchmarks}
The self-contained proof workflow takes approximately 2--4 hours, and the full $k=1024,d=5$ registry-growth workload takes approximately 2--3 hours on the reference desktop. Together they remain below one day. Mobile, external-artifact, service, and Linux-netem workflows are retained or optional because they require named devices, network downloads, containers, or specialized network control.

\subsection{Artifact Installation \& Configuration}
From a fresh checkout, reviewers run:
\begin{verbatim}
$ ./artifact.sh doctor
$ ./artifact.sh verify
$ ./artifact.sh smoke
\end{verbatim}
These commands check the toolchain and release manifest, build and test both implementations, exercise the composed credential/status relation, and run a one-sample SMT control. The smoke test is functional rather than timed, provides paper evidence, and ends with \texttt{SMOKE PASS}.

\subsection{Major Claims}

\begin{itemize}
    \item[(C1)] The Verkle relation accepts five completeness cases and rejects thirteen selected invalid statements or witnesses (Section~\ref{sec:experiments}; E1 and E2).
    \item[(C2)] Experiment E2 supports Section~\ref{sec:experiments} and Table~\ref{tab:proof-side-costs} by reproducing proof and verification costs for both proof-system backends for \textsc{ShadowPath}-Verkle, the address-space-matched SMT control, authenticated state reconstruction and updates, and executable credential/status composition.
    \item[(C3)] Experiments E3 and E4 support Section~\ref{sec:experiments} and Figures~\ref{fig:registry_growth_comm}, \ref{fig:mobile_soc_pareto}, and \ref{fig:comparison}: the full registry-growth workload is reproducible, and retained or optional evidence covers IPFS/IPNS synchronization, verifier concurrency, external-system context, and physical-device results.
\end{itemize}

\subsection{Evaluation}

\subsubsection{Experiment (E1): Functional Validation}

\textit{[Time]} About 30--60 minutes on a fresh host; warm caches shorten the run.

\textit{[Preparation]} In an unmodified artifact checkout, run the \texttt{doctor} and \texttt{verify} subcommands.

\textit{[Execution]} Run \texttt{./artifact.sh smoke}.

\textit{[Results]} The command ends with \texttt{SMOKE PASS}, exercises both implementations, and supports functional inspection of C1; its one-sample timing is not paper evidence. The retained suite has five accepting and thirteen selected rejecting cases.

\subsubsection{Experiment (E2): Desktop Reproduction}

\textit{[Time]} About 2--4 compute-hours and 10 human-minutes.

\textit{[Preparation]} Reserve 5~GB and close competing workloads.

\textit{[Execution]} Run \texttt{./artifact.sh reproduce}. If export or plot generation is interrupted, resume with \texttt{./artifact.sh finalize}.

\textit{[Results]} Five completeness cases are accepted, and 13 selected invalid cases are rejected; all 24 authenticated state configurations reproduce their expected roots. Reference proving/verification medians are 2109.52/7.553~ms with a 484-byte proof for Verkle Groth16, 25619.79/9.595~ms with a 1352-byte proof for Verkle PLONK, and 371.645/3.703~ms with a 388-byte proof for SMT Groth16. Proof sizes and logical outcomes must match exactly. Absolute latency is host-dependent and has no cross-host percentage bound; the expected same-host ordering is SMT Groth16, Verkle Groth16, then Verkle PLONK by proving latency. This experiment supports C1 and C2.

\subsubsection{Experiment (E3): Registry Growth}

\textit{[Time]} About 2--3 compute-hours and 5 human-minutes.

\textit{[Preparation]} Close competing workloads and reserve 5~GB of free disk space.

\textit{[Execution]} Run \texttt{./artifact.sh registry}. The scaled \texttt{REGISTRY\_COUNTS=100,1000} form is a kick-the-tires workflow check.

\textit{[Results]} The full command evaluates 100, 1,000, 10,000, and 100,000 revoked entries at $k=1024,d=5$. Complete-summary size grows from approximately 0.118~MB to 93.95~MB in decimal units. The result supports C3 and Figure~\ref{fig:registry_growth_comm}; timing remains host-dependent.

\subsubsection{Experiment (E4): Retained and Optional Evidence}

\textit{[Time]} Plot validation takes about 5 minutes; \texttt{README.md} lists times for optional external, Docker, and device workflows.

\textit{[Preparation]} Consult \texttt{README.md} for Docker, Linux-netem, external-artifact, and mobile prerequisites; supplied mobile exports can be validated unchanged.

\textit{[Execution]} Run \texttt{./artifact.sh plots} to regenerate retained-CSV plots or \texttt{./artifact.sh mobile} to re-import retained device exports into a separate tree; \texttt{README.md} covers optional Docker and external artifacts.

\textit{[Results]} These workflows validate retained synchronization, cross-system, and device evidence supporting C3. They remain outside the requested desktop Reproduced scope.

\medskip
\noindent\textbf{Physical-device rerun (optional).} The shared Go bridge is in \texttt{go-verkle-zk/mobilebench}; \texttt{ios\_evaluator} and \texttt{android\_evaluator} contain the platform apps and detailed READMEs. Both apps must report circuit identifier \texttt{shadowpath-mode-b-hardened-v3}.

\textit{[iOS build]} On macOS with Go~1.25, Xcode~16, XcodeGen, and an Apple signing team, run:
\begin{verbatim}
$ cd ios_evaluator
$ bash build_xcframework.sh
$ xcodegen generate
$ open ZKPEvaluator.xcodeproj
\end{verbatim}
Select the \texttt{ZKPEvaluator} scheme, configure local signing, and install a Release build on a physical iOS~16-or-newer device.

\textit{[Android build]} With Go~1.25, JDK~17, Android SDK~35, and NDK~\texttt{30.0.14904198}, run:
\begin{verbatim}
$ cd android_evaluator
$ bash build_aar.sh
$ ./gradlew assembleDebug
$ cd app/build/outputs/apk/debug
$ adb install -r app-debug.apk
\end{verbatim}
This build checks functionality; use an Android Studio Release build without an attached debugger for timed measurements on a physical device running Android 7 or newer.

\textit{[Measurement]} Uninstall older builds, install the current Release build, reboot the phone, connect power, disable Low Power/Battery Saver, close other apps, and keep the evaluator in the foreground. Without a bundled seed cache, the first launch generates and persists public proving parameters before the timed loops; the platform READMEs describe optional host-side seeding. In the app, select \texttt{Bench (5 iters)} and \texttt{Full benchmark}. This runs PLONK--Groth16 and Groth16--PLONK with five trials per backend/order and a 60-second cooldown. Export the dated \texttt{trials.csv}, \texttt{results.json}, \texttt{telemetry.csv}, and \texttt{log.txt} files from \texttt{Records}; preserve them unchanged and confirm both key-set digests and the circuit identifier. From the repository root, \texttt{./artifact.sh mobile} validates retained exports and regenerates normalized CSV and \LaTeX{} outputs. Simulator and emulator results are functional checks, not paper evidence.

\end{document}